\documentclass{report}

\usepackage{techreport}

\usepackage{graphicx}
\usepackage{booktabs}
\usepackage{array}
\usepackage{amsmath}
\usepackage{amssymb}
\usepackage{amsfonts}
\usepackage{multirow}
\usepackage{verbatim}
\usepackage{caption}
\usepackage{longtable}
\usepackage{supertabular}
\usepackage{float}
\usepackage{enumitem}
\usepackage{tablefootnote}
\usepackage{xcolor}
\usepackage{colortbl}
\usepackage{xspace}
\usepackage{textcomp}
\usepackage{makecell}
\usepackage{todonotes}
\usepackage{listings}
\usepackage{soul}
\usepackage{lscape}
\usepackage{siunitx}

\usepackage{pifont}%
\usepackage{scrextend}

\usepackage{array}
\usepackage{tgpagella}
\usepackage{latexsym}
\usepackage[T1]{fontenc}
\usepackage[utf8]{inputenc}
\usepackage{microtype}
\definecolor{mydarkblue}{rgb}{0,0.08,0.45}
\definecolor{mylightblue}{RGB}{39,114,191}
\usepackage[colorlinks,citecolor=mylightblue,urlcolor=mylightblue,linkcolor=mylightblue]{hyperref}
\usepackage{url}            %
\usepackage{nicefrac}       %
\usepackage{changepage}
\usepackage{xargs}          %
\usepackage{wrapfig,lipsum,booktabs}
\usepackage{longtable}
\usepackage{subcaption}
\usepackage{endnotes}
\usepackage{footnote}

\usepackage{pgfplots}
\usetikzlibrary{pgfplots.groupplots}
\pgfplotsset{compat=1.3}

\usepackage[most]{tcolorbox}

\usepackage[capitalize,noabbrev]{cleveref}

\usepackage{multicol}
\usepackage{multirow}

\usepackage{tcolorbox}
\usepackage{titlesec}

\definecolor{myyellowgreen}{RGB}{154, 205, 50}

\usepackage{minitoc}

\newcommand{\IGNORE}[1]{}
\usepackage{algorithm}
\usepackage{algorithmicx}
\usepackage{algpseudocode}

\usepackage[title]{appendix}

\newcommand\simplefootnote[1]{%
  \begingroup
  \renewcommand\thefootnote{}\footnote{#1}%
  \addtocounter{footnote}{-1}%
  \endgroup
}

\newcommand{\msrvshort}[1]{MSRV}
\newcommand{\msrvlong}[1]{Microsoft Research Vancouver}

\title{\vspace{-2em}%
  \hrule height 4pt%
  \vskip 0.25in%
  \vskip -\parskip%
  \textbf{
FengHuang: Next-Generation Memory Orchestration for AI Inferencing
}

  \vskip 0.2in%
  \vskip -\parskip%
  \hrule height 1pt%
  \vskip 0.09in}

\author{
Jiamin Li $^*$ \hspace{0.12cm}
Lei Qu $^*$ \hspace{0.12cm}
Tao Zhang $^*$ \hspace{0.12cm}
Grigory Chirkov $^*$ \hspace{0.12cm}
Shuotao Xu $^*$ \hspace{0.12cm}
Peng Cheng \hspace{0.12cm} 
Lidong Zhou \vspace{0.2cm} \\ 
\textbf{\normalsize Microsoft Research} \vspace{0.4cm} \\
\{jiaminli, lequ, zhangt, gchirkov, shuotaoxu, pengc, lidongz\}@microsoft.com \vspace{0.4cm} \\
November, 2025
}

\newcommand{\phoenix}{\textsc{FengHuang}\xspace}
\newcommand{\localmem}{\textit{xPU Local Memory}\xspace}

\newcommand{\remotemem}{\phoenix \textit{Remote Memory}\xspace}
\newcommand{\tabfull}{\phoenix \textit{Tensor Addressable Bridge}\xspace}
\newcommand{\tab}{\textit{TAB}\xspace}
\newcommand{\prefetcher}{\textit{Tensor Prefetcher}\xspace}

\begin{document}
\maketitle

\simplefootnote{* All authors contributed equally to this report.}

\begin{abstract}
\emph{This document presents a vision for a novel AI infrastructure design that has been initially validated through inference simulations on state-of-the-art large language models.}

Advancements in deep learning and specialized hardware have driven the rapid growth of large language models (LLMs) and generative AI systems. 
However, traditional GPU-centric architectures face scalability challenges for inference workloads due to limitations in memory capacity, bandwidth, and interconnect scaling. 
To address these issues, the \textbf{\phoenix Platform}, a disaggregated AI infrastructure platform, is proposed to overcome memory and communication scaling limits for AI inference. 
\phoenix features a multi-tier shared-memory architecture combining high-speed local memory with centralized disaggregated remote memory, enhanced by active tensor paging and near-memory compute for tensor operations. 
Simulations demonstrate that \phoenix achieves \textbf{up to 93\% local memory capacity reduction, 50\% GPU compute savings, and 16$\times$ to 70$\times$ faster inter-GPU communication} compared to conventional GPU scaling. 
Across workloads such as GPT-3, Grok-1, and QWEN3-235B, \phoenix enables \textbf{up to 50\% GPU reductions} while maintaining end-user performance, offering a scalable, flexible, and cost-effective solution for AI inference infrastructure. 
\phoenix provides an optimal balance as a rack-level AI infrastructure scale-up solution. 
Its open, heterogeneous design eliminates vendor lock-in and enhances supply chain flexibility, enabling significant infrastructure and power cost reductions.
\end{abstract}

\clearpage
\tableofcontents
\clearpage

\chapter{Introduction}


Ever since the first public release of ChatGPT service in 2022, the consumption of various LLM-based AI services has been growing at a breakneck pace~\cite{explodingtopics2023}. 
Sources estimate that the number of different AI tool users worldwide has increased from 116 million people in 2020 to 314 million people in 2024, an almost threefold increase in the span of just 4 years~\cite{altindex2025}. 
Naturally, this results in growing demand for AI hardware (GPUs, TPUs, etc.) and increased datacenter utilization.

The continuing scaling of the models exacerbates this situation even further. 
New generations of foundation models often reach scales that dwarf their predecessors. 
For example, various analyses and reports estimate that the GPT models grew from around 175 billion parameters (GPT-3, released in 2020) to approximately 1.8 trillion parameters (GPT-4, released in 2023) in just 3 years~\cite{brown2020gpt3, thedecoder2023gpt4}. 
Needless to say, this factor worsens the strain on AI datacenter infrastructure even further. 

\begin{figure}[h]
\centering
\includegraphics[width=0.95\linewidth]{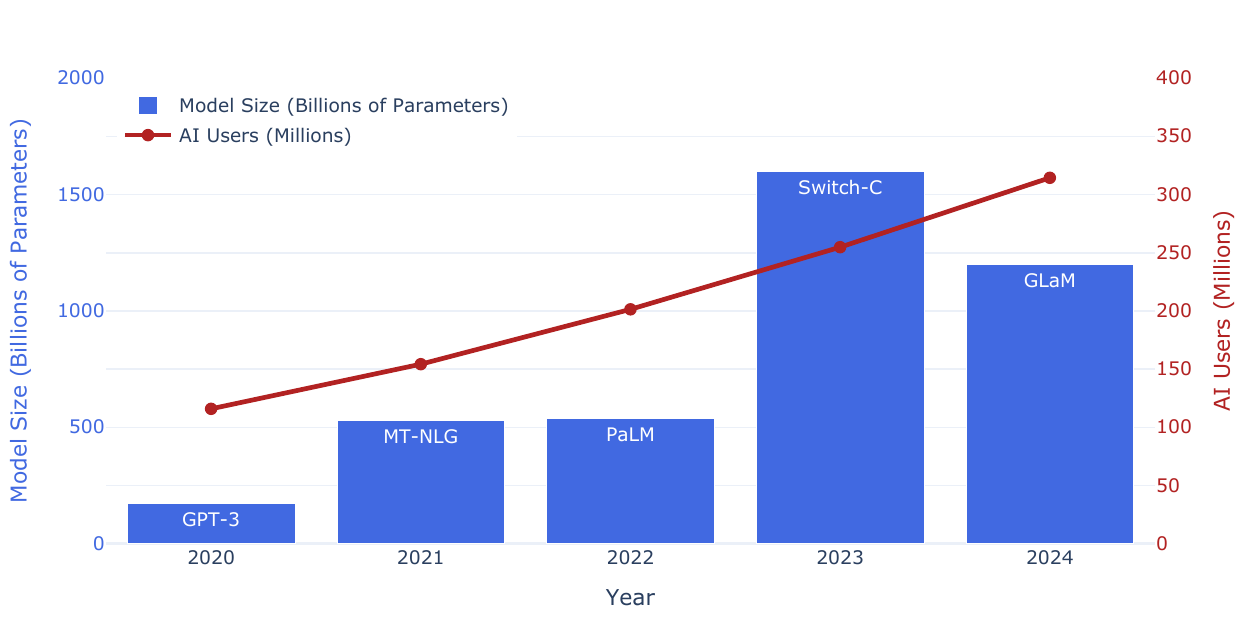}
\caption{Growing number of AI users worldwide and growing sizes of the state-of-the-art AI models. 
Data is taken from from~\cite{brown2020gpt3,smith2021mtnlg,chowdhery2022palm,fedus2022switch,du2022glam,resourcera2025,altindex2025}.
}
\label{fig:ai-trends}
\end{figure}

Fig.~\ref{fig:ai-trends} shows how two previously mentioned metrics developed over time. Together, these two factors has led to unprecedented expenses from the leading Cloud providers, including both Capital Expenses (CapEx) and Operational Expenses (OpEx). 
On the CapEx side specifically, hyperscalers are doing their best to capture the endlessly growing demand for AI hardware caused by the popularity of the AI tools. 
Looking at the investments in AI infrastructure from large datacenter companies in their fiscal years 2025 alone, Microsoft allocated over \$80 billion, Meta is spending \$65 billion, Google is investing \$75 billion, and Amazon is committing \$100 billion~\cite{microsoft_ai_investment_2025, google_ai_investment_2025, amazon_ai_investment_2025, meta_ai_investment_2025}. 


In light of things mentioned above, it becomes clear that optimizing the efficiency and the cost of AI infrastructure is a crucial prerequisite for further scaling of AI models and make AI tools more accessible. 
One of the prime targets ripe for rethinking and reimagining for contemporary AI workloads is the \textbf{memory subsystems} in the modern AI accelerators.

Indeed, in the last couple of years the evolution of AI models has yielded \textbf{dissimilarity} in scaling trends of workloads' compute volume, memory footprint, memory usage per invocation, and inter-device communication.
Specifically, our data in Chapter~\ref{chap:background} shows that state-of-the-art models are increasingly less constrained by hardware compute throughput with each generation.
Instead, newer models tend to be increasingly starved for memory throughput, memory capacity, and inter-device communication throughput. 

\begin{figure}[h]
\centering
\includegraphics[width=0.95\linewidth]{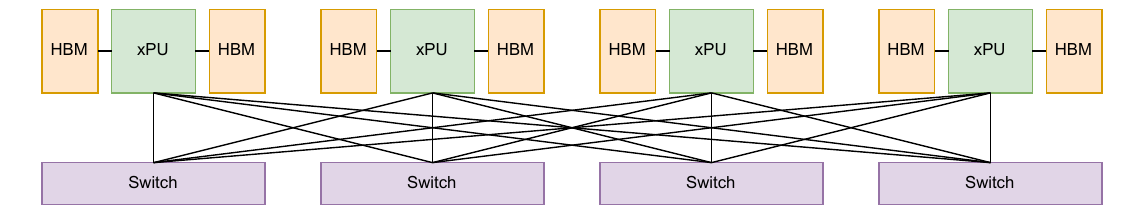}
\caption{
    High-level design of a conventional inference node.
    Separate devices are bound by inter-device interconnect with relatively low bandwidth.
    In such systems, increasing memory capacity necessarily entails adding additional compute hardware.
    Sharing data between xPUs involves transferring data over a relatively slow interconnect.
}
\label{fig:conv_inf_node}
\end{figure}

In contrast, the latest generations of AI accelerators from the leading vendors (e.g., NVIDIA~\cite{nvidia_h100_whitepaper_2022}, AMD~\cite{amd_mi300x_whitepaper_2025}, and Google~\cite{jouppi2023tpuv4}, etc.) tend to increase their compute throughput faster than other metrics with each generation. 
Moreover, the proposed memory capacity scaling mechanism is essentially joining multiple accelerators into a single node using relatively low-bandwidth (compared to HBM) interconnect.
Fig.~\ref{fig:conv_inf_node} shows such design.
As a result, datacenter providers are forced to scale memory capacities in conjunction with compute throughput.
While this enables running large state-of-the-art models, it leaves accelerators’ execution units severely underutilized.
Moreover, accessing memory over inter-xPU interconnect creates a bottleneck, and generally discourages too much data sharing between chips.

This technical report presents the findings of \textit{Project \phoenix}: a Microsoft Research initiative.
The project's main goal is to \textbf{reduce the cost of AI inference infrastructure by redesigning accelerator memory systems without sacrificing the performance}.
The result of this year-long project is the AI node design that completely rethinks the memory subsystems of AI accelerators.

\begin{figure}[h]
\centering
\includegraphics[width=0.95\linewidth]{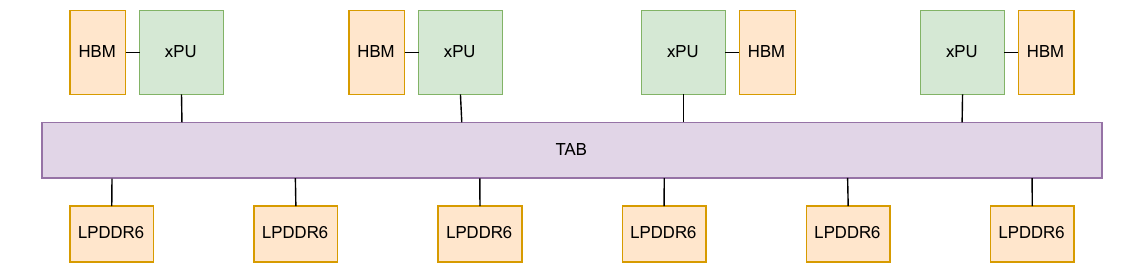}
\caption{
    High-level design of a \phoenix node.
    Most of the node's memory is connected to the Tensor Addressable Bridge (TAB) and is shared between all xPUs.
    Memory scaling can be done separately from compute scaling.
    Exchanging data between devices is quick and done through the shared memory.
}
\label{fig:phoenix_high_level}
\end{figure}

Fig.~\ref{fig:phoenix_high_level} shows this design.
At a high level, the new architecture adds a new LPDDR6 memory level that is shared between all the accelerators (xPUs).
Specifically, it introduces new memory space that is sharded across all LPDDR6 modules and is accessible by all xPUs.
It is meant to be a high-capacity (order of Terabytes) volume that stores the model's weights and intermediate results that are not used immediately. 
Using LPDDR6 memory instead of conventional HBM modules gives the advantage of much higher capacity while meeting the high access bandwidth requirements of the LLM inference workloads.
Tensor Addressable Bridge (TAB) is a novel chip that connects xPUs to each other and to the new memory modules.
The interfaces between TAB and xPUs are based on modern SerDes transceivers operating at 224G/448G speeds~\cite{224gserdes} and enables TAB memory to operate at roughly the same bandwidth as conventional tightly-coupled HBM memory.
A combination of novel hardware tensor prefetchers in xPUs and software modifications implements the paging algorithm that brings the tensors from the TAB memory into the local xPU memories just in time before usage in computation and evicts the newly produced tensors back into the TAB memory when needed.
Data exchange between xPUs is achieved through read and write operations to the shared memory.
Additional computational hardware in TAB supports line-rate in-memory tensor reduction.

\phoenix architecture helps tackle two of the previously mentioned inefficiencies.
First, it allows tuning the memory capacity of the node separately from its computational throughput.
Each new generation of TAB can be designed to support either more computational throughput, or more memory capacity.
Moreover, \phoenix's scaling mechanism allows increasing memory capacity by connecting more than one TAB to each GPU, thus decoupling memory capacity from computational throughput.
This will help balance these two resources, avoid xPU underutilization, and decrease the costs of the hardware.
Second, the new shared memory approach eliminates the interconnect bottleneck and makes data transfers between xPUs virtually free.
In contrast to previous approach, this design encourages more data sharing between the xPUs.

We evaluate \phoenix by comparing an 8-GPU H200 node with a \phoenix system containing four similar GPUs connected through one TAB.
Our evaluation shows that \phoenix system provides memory capacity scalability and better inter-xPU communication while achieving similar or better performance in both prefill and decode phases when running inference using the state-of-the-art LLM models.
This shows that \phoenix can reduce the xPU computation chips by half while maintaining AI inference workload performance, thus resulting in significantly lower AI infrastructure cost and operational expenses compared to existing monolithic xPU architecture. 

The design shown in this report uses only very generic assumptions about the xPU architecture and can be readily adapted to all modern AI accelerators.
We strongly believe that the solution shown in this paper is superior to the architectures that are currently deployed in the world's AI datacenters and should be adopted to improve inference efficiency and enable the deployment of the new, more capable AI models.

\chapter{Background and Motivation}
\label{chap:background}
\section{Model vs. Hardware: Trends}
\label{sec:model_vs_hardware}
LLM architectures are evolving at a remarkable pace. This rapid progress is driven by the pursuit of improved performance, accuracy, and the capability to handle complex data and reasoning tasks. These advancements are crucial for applications such as automated reasoning and agentic AI.
In this section, we analyze both the workload and hardware evolution trends to identify where hardware lags behind workload requirements.

\subsection{Memory Capacity Wall}
\label{sec:computation_hbm_cap}
The evolution of LLM architectures places significant pressure on the memory capacity of accelerators in inference scenarios.
This growing demand for greater memory capacity is driven by key factors such as parameter scaling, long-context support, and system efficiency requirements for large inference batch sizes.
\subsubsection{Inference Workload Trend 1 -- Increasing Memory Capacity Requirement}
\begin{figure}[h]
\centering
\includegraphics[width=0.6\linewidth]{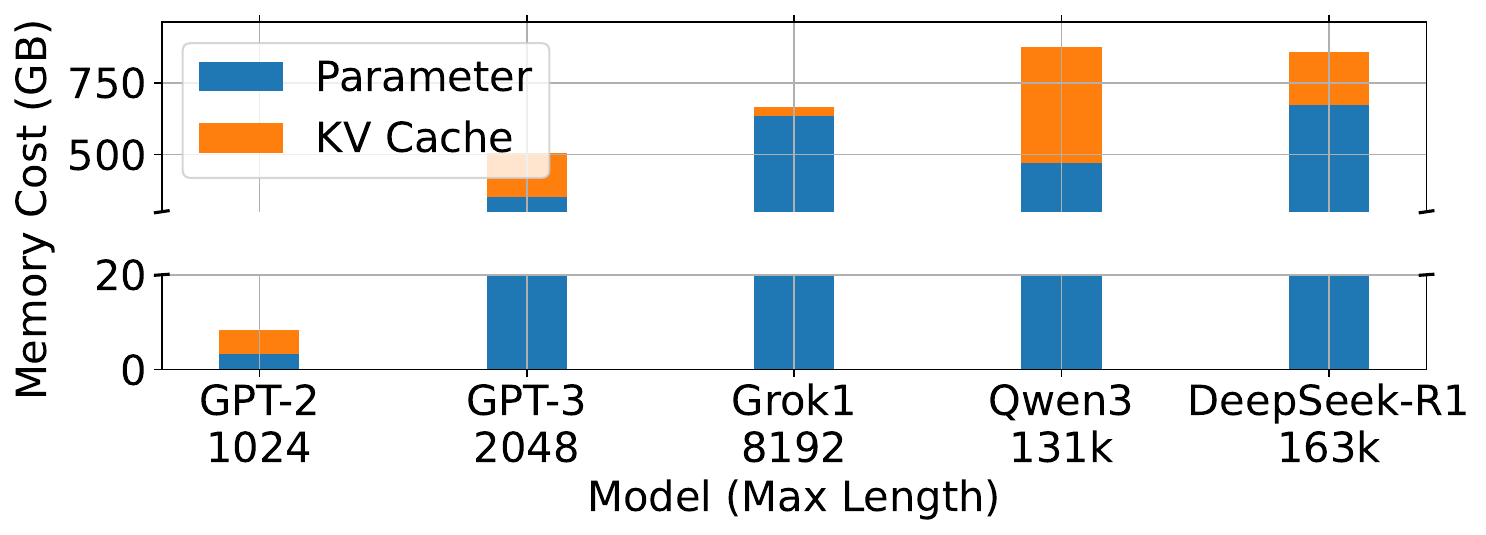}
\caption{Model Memory Capacity Requirements (Batch Size = 16)}
\label{fig:model_trend_memory}
\end{figure}

LLM inference workloads rely heavily on xPUs’ memory capacity to accommodate two key data structures: model parameters and the KV-Cache.
Both components are integral to the model’s performance but increasingly strain hardware resources as model scales grow (Figure~\ref{fig:model_trend_memory}).

\begin{figure}[h]
\centering
\includegraphics[width=0.6\linewidth]{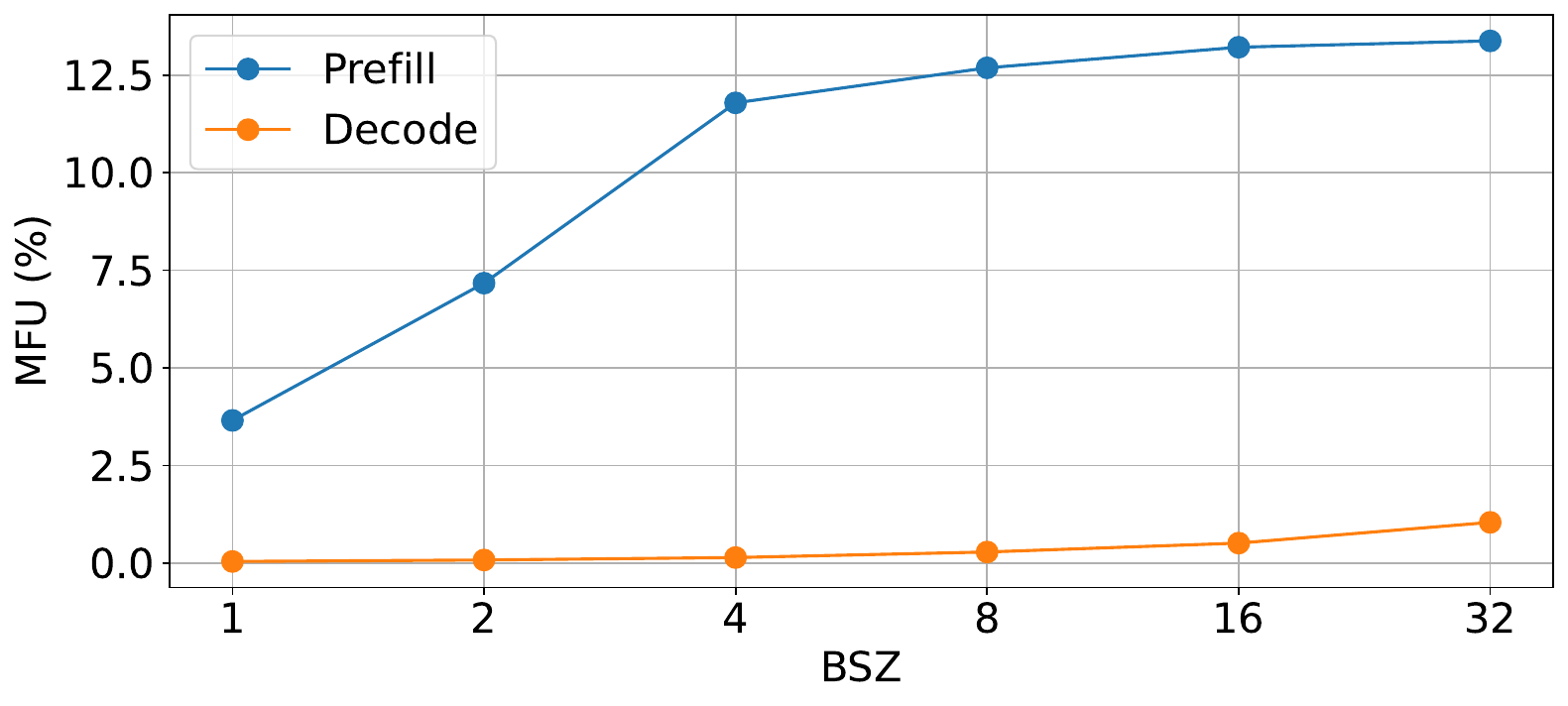}
\caption{MFU (Model FLOPs Utilization) vs. Batch Size}
\label{fig:mfu_vs_batch}
\end{figure}

\begin{itemize}
\item\textbf{Model Parameter Size Scaling}

LLM development exhibits a clear trend of increasing parameter counts, which enhance model intelligence but also impose significant memory demands.
For example, the transition from GPT-3 (175B parameters) to DeepSeek-V3 (671B parameters) represents approximately a 3.8$\times$ increase in size, with a 671B-parameter model in FP16 requiring over 1.34 TB of HBM.
\textbf{Low-bit precision techniques}, such as FP8 in DeepSeek-V3, reduce storage requirements by roughly half; however, DeepSeek-V3 still demands nearly twice the memory of GPT-3, highlighting the persistent challenge of scaling model capacity while optimizing memory efficiency.

\item\textbf{Model KV-Cache Size Scaling}
\begin{itemize}
\item\textbf{Sequence-length scaling:}
As models advance to handle more complex reasoning, the context length during inference grows significantly, driving up KV-Cache memory requirements.
For instance, reasoning models such as Qwen3 and DeepSeek require maximum sequence lengths of 128K and 160K, far surpassing Grok1’s 8K.
Techniques like MLA (Multi-head Latent Attention) compression, used in DeepSeek, reduce KV-Cache footprints by up to 10$\times$ compared to conventional MHA.
However, as agentic AI applications gain traction, the demand for long-context reasoning is expected to increase further, raising both the maximum sequence length and KV-Cache memory requirements.
\item\textbf{Batch-size scaling:}
KV-Cache size scales linearly with batch size, making large batches a primary driver of memory demand.
GPUs, optimized for throughput, rely on large batches to maximize FLOPs utilization and overall hardware efficiency.
As shown in Figure~\ref{fig:mfu_vs_batch}, MFU increases significantly with batch size, enhancing computational utilization.
Large-scale inference systems such as SGLang adopt this strategy by minimizing memory allocated to model parameters while reserving most GPU memory for KV-Cache storage—enabling efficient inference at scale despite growing KV-Cache size demands.
\end{itemize}
\end{itemize}

\subsubsection{Inference Workload Trend 2 -- Steady Computational Requirement}
\begin{figure}[h]
\centering
\includegraphics[width=0.6\linewidth]{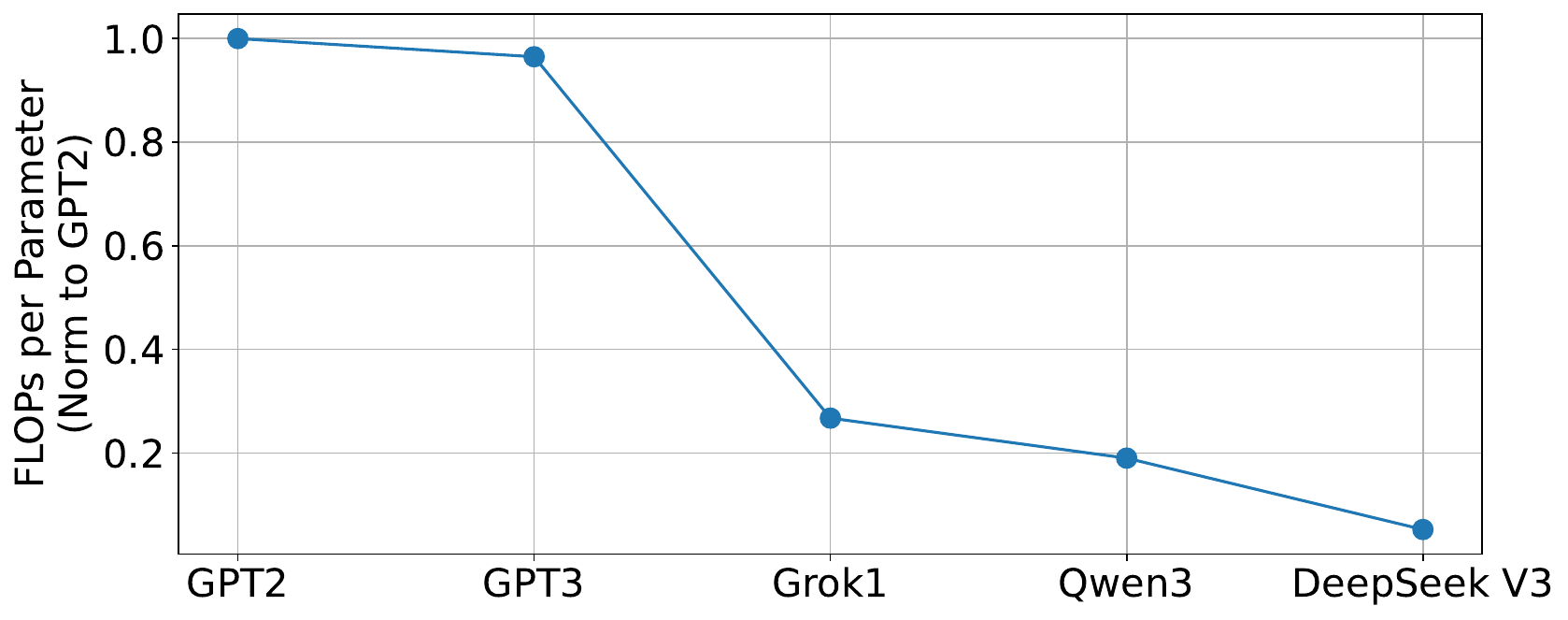}
\caption{Trends of Model FLOPs per generated token requirements (1K KV-Cache)}
\label{fig:model_trend_flops}
\end{figure}

Despite the significant growth in the model’s memory requirements, the number of FLOPs required to generate a single token remains steady or even decreases.
As shown in Figure~\ref{fig:model_trend_flops}, computational demands increase sharply between GPT-2 and GPT-3 due to their dense architectures but subsequently stabilize or decline.
The introduction of Sparse Mixture-of-Experts (MoE) models has played a key role in this trend, as they activate only a small subset of parameters during inference.
For example, models such as DeepSeek-V3 leave up to 95\% of parameters inactive during inference, thereby preventing computational load from scaling with model size and enabling larger models without longer inference time.

\subsubsection{Hardware vs. Workload Trend -- Diverging Scaling of Hardware FLOPs and HBM Capacity from Workload Requirements}
\begin{figure}[h]
\centering
\includegraphics[width=0.6\linewidth]{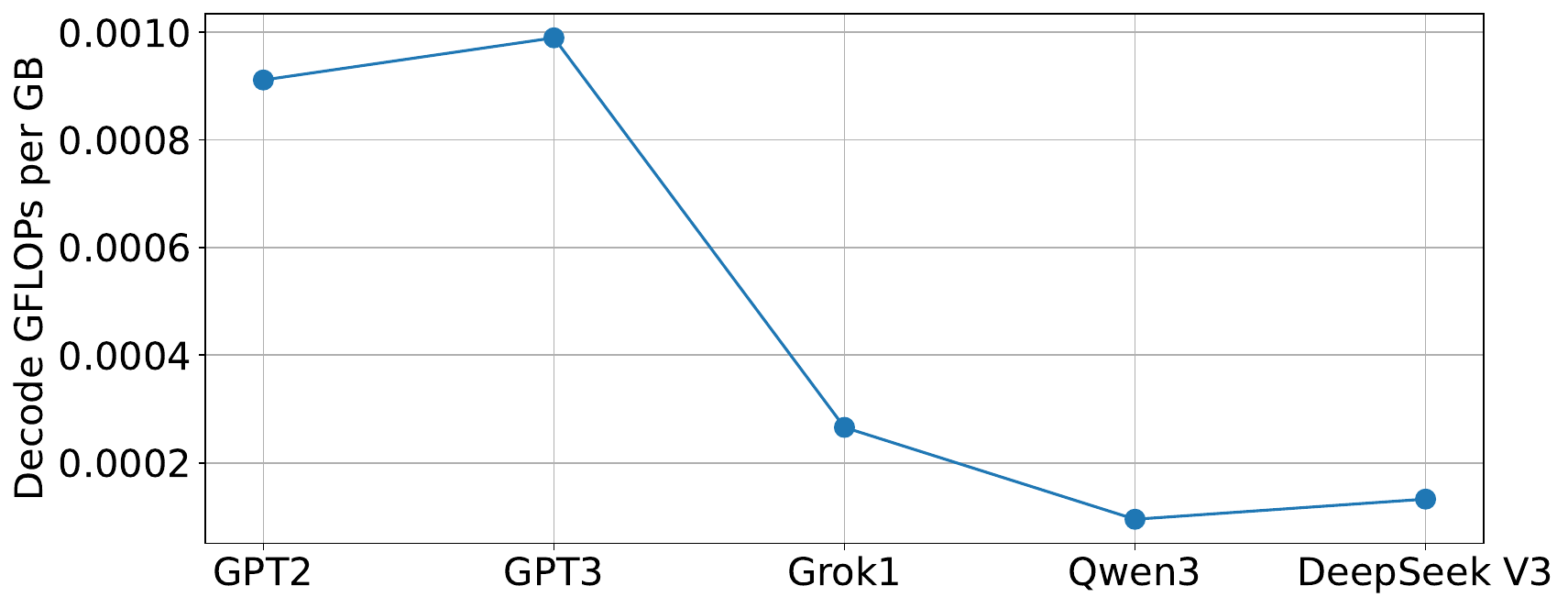}
\caption{Trends of model computation and memory capacity requirements ratio.}
\label{fig:flop_param_trend}
\centering
\includegraphics[width=0.6\linewidth]{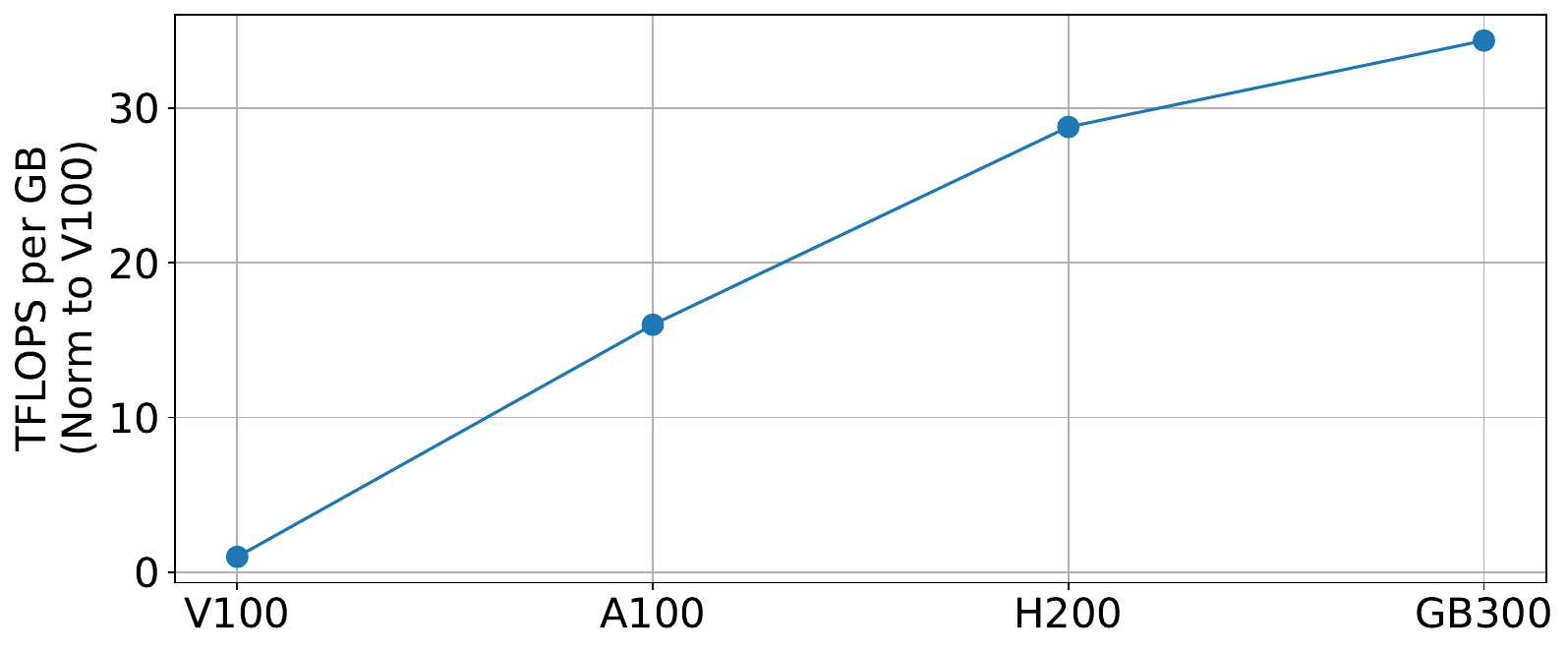}
\caption{Trends of hardware computation (FLOPS) and memory capacity (GB) ratio.}
\label{fig:flop_cap_hardware}
\end{figure}

In the evolving landscape of large language models (LLMs), model sizes continue to grow, while per-token computational requirements (FLOPs per generated token) have remained relatively stable due to algorithmic innovations such as Sparse Mixture-of-Experts (MoE).
As illustrated in Figure~\ref{fig:flop_param_trend}, the ratio of computational operations per generated token to the model’s memory footprint has decreased by roughly an order of magnitude from GPT-2 to DeepSeek-V3.
However, hardware development trends show an opposite trajectory.
As shown in Figure~\ref{fig:flop_cap_hardware}, the ratio of computational capability to memory capacity in GPUs has increased significantly, diverging from the scaling patterns observed in LLMs.
While xPU computational performance has advanced steadily, memory capacity scaling—i.e., high-bandwidth memory (HBM) capacity—has lagged behind.
For instance, the FLOPs-per-GB-capacity ratio of GPUs has risen by approximately 34$\times$ from the V100 to the GB200 architecture.
This growing mismatch between hardware scaling and model requirements poses a critical challenge.
The continued emphasis on computational throughput over memory capacity risks becoming a major impediment to achieving efficient AI infrastructure for inference workloads, ultimately constraining the scalability and performance of future AI systems.

\subsection{Memory Bandwidth Wall}
\label{sec:memory_bandwidth_wall}
As LLMs continue to scale, memory bandwidth has emerged as a critical bottleneck in achieving high inference efficiency.
While xPUs are optimized for computational throughput, LLM inference workloads often require substantially higher memory bandwidth per unit of computation.
This mismatch underscores the growing need for hardware capable of sustaining memory-intensive operations effectively.
\subsubsection{Model Requirement: High Byte-per-FLOP Ratio Requires High Memory Bandwidth}
LLM inference workloads exhibit distinct characteristics in two phases — prefill and decode.
The prefill phase is generally compute-intensive, although increasing model sparsity has been shifting prefill operations into memory-bound regions (\cref{fig:prefill_decode_hardware}).
In contrast, the decode phase is predominantly memory-intensive, requiring frequent access to large volumes of data while performing relatively light computation.
To quantify this behavior, we define the Byte-per-FLOP ratio as a measure of the amount of data accessed relative to the number of FLOPs executed.
For example, in Qwen3, the Byte-per-FLOP ratio during the decode phase is roughly 100$\times$ higher than during prefill (\cref{fig:prefill_decode_hardware}), highlighting the extreme memory bandwidth demands of decode operations.
\subsubsection{Hardware Limitation: Low Byte-per-FLOP in GPUs}
\begin{figure}[h]
    \centering
    \includegraphics[width=0.6\linewidth]{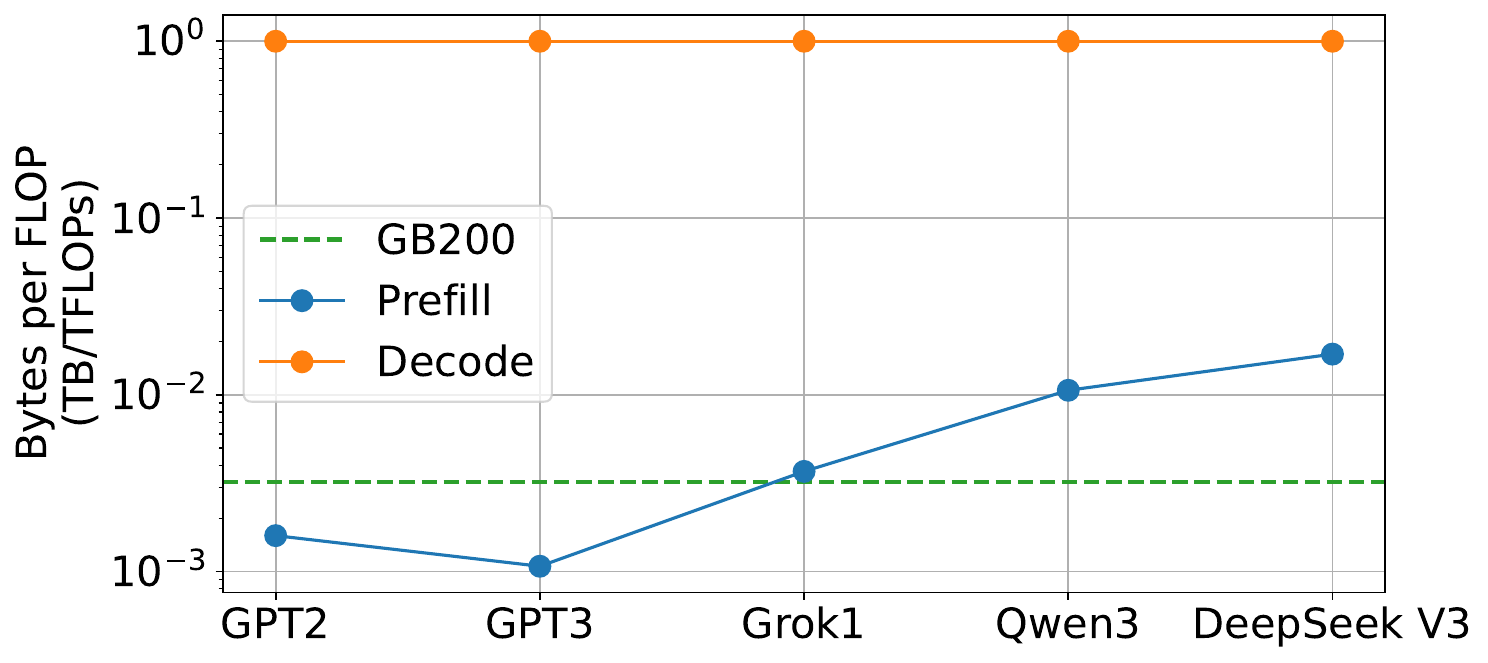}
    \caption{Ratio between memory traffic and FLOPs in various LLM model's prefill and decode and between GB200's memory bandwidth and FLOPS.}
    \label{fig:prefill_decode_hardware}
\end{figure}

\begin{figure}[h]
    \centering
    \includegraphics[width=0.7\linewidth]{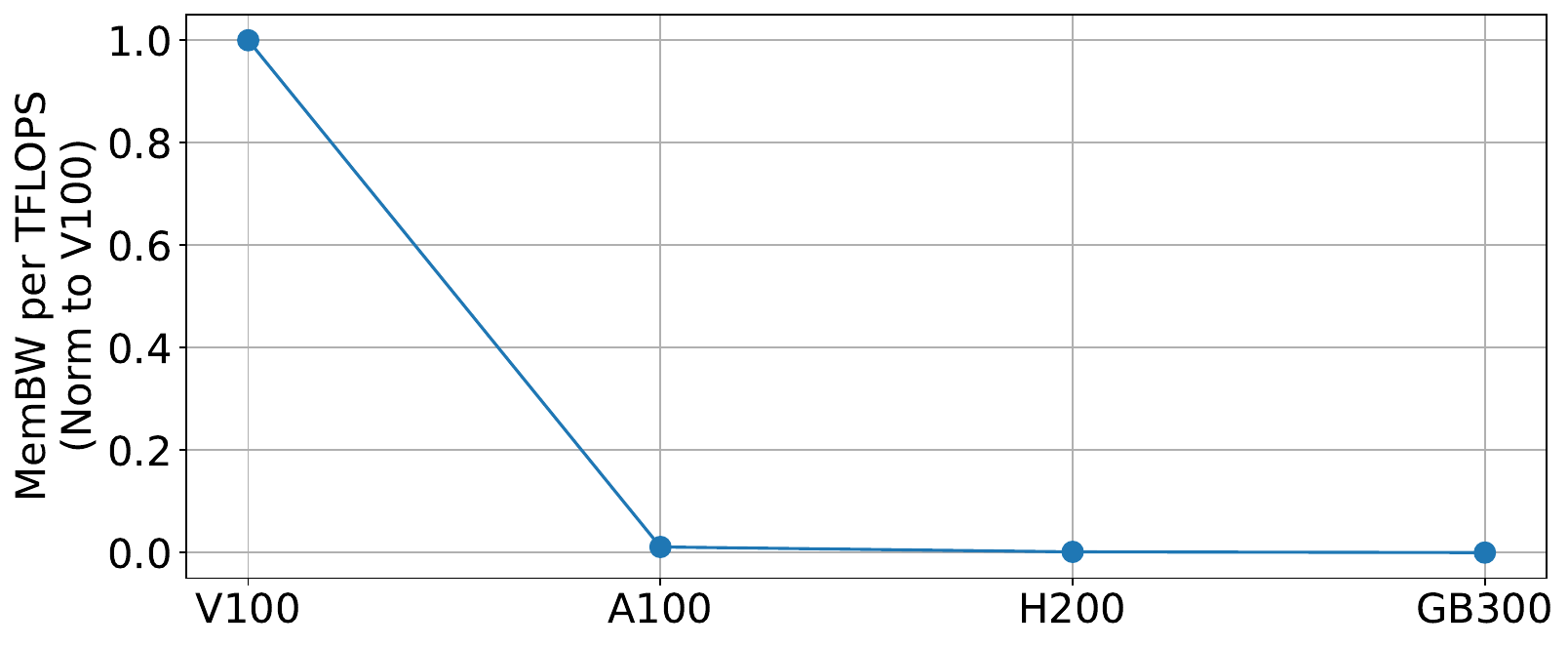}
    \caption{Trend of ratio between memory bandwidth and FP16 FLOPS of recent xPUs.}
    \label{fig:hbmbw_flop_ratio_hardware}
\end{figure}

Existing xPU architectures are designed for compute-intensive workloads with high FLOPs-per-second capabilities.
However, these xPUs typically exhibit a low Byte-per-FLOP ratio, indicating insufficient memory bandwidth relative to their computational power.
This limitation has become increasingly pronounced across successive GPU generations, as shown in \cref{fig:hbmbw_flop_ratio_hardware}.
The impact of a low Byte-per-FLOP ratio is particularly evident during the memory-intensive decode phase.
Comparing the Byte-per-FLOP ratios of models’ prefill and decode phases with GB300’s hardware ratio in \cref{fig:prefill_decode_hardware} and \cref{fig:hbmbw_flop_ratio_hardware} reveals a clear and widening gap between model demands and existing hardware capabilities.


\subsection{Inter-Device Interconnect Bandwidth Wall}
When the number of model parameters exceeds the capacity of a single xPU, multi-xPU orchestration becomes necessary to support distributed inference.
In such cases, the inter-xPU interconnect is heavily utilized to aggregate and exchange intermediate results among multiple devices.
\subsubsection{Model Width and Communication Volume}
In distributed inference, frequent and large-scale data exchanges occur between xPUs.
The volume of transferred data is primarily determined by the model’s hidden size: a larger hidden size generally results in greater data transfer per communication step.
As shown in Figure~\ref{fig:model_flop_comm}, the five models have hidden sizes of 768, 2048, 6144, 5120, and 7168, respectively, and their transferred data volumes follow a similar pattern.
Although Grok1, Qwen3, and DeepSeek-V3 possess comparable hidden sizes, the sparse MoE architectures used in Qwen3 and DeepSeek-V3 yield significantly lower FLOPs per transfer byte compared to Grok1.
\subsubsection{NVLink Bandwidth Scaling Lagging Behind Compute}
\begin{figure}[h]
\centering
\includegraphics[width=0.6\linewidth]{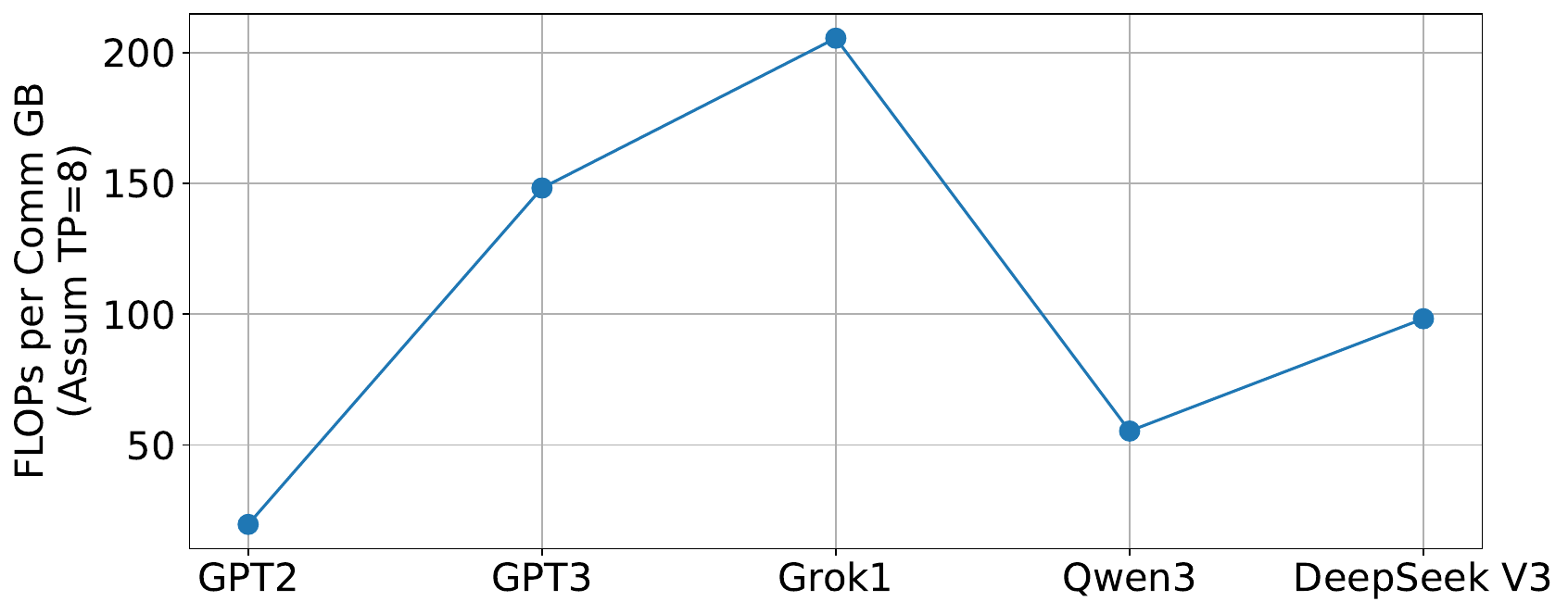}
\caption{Trends of ratio between model's FLOPs and communication sizes.}
\label{fig:model_flop_comm}
\centering
\includegraphics[width=0.6\linewidth]{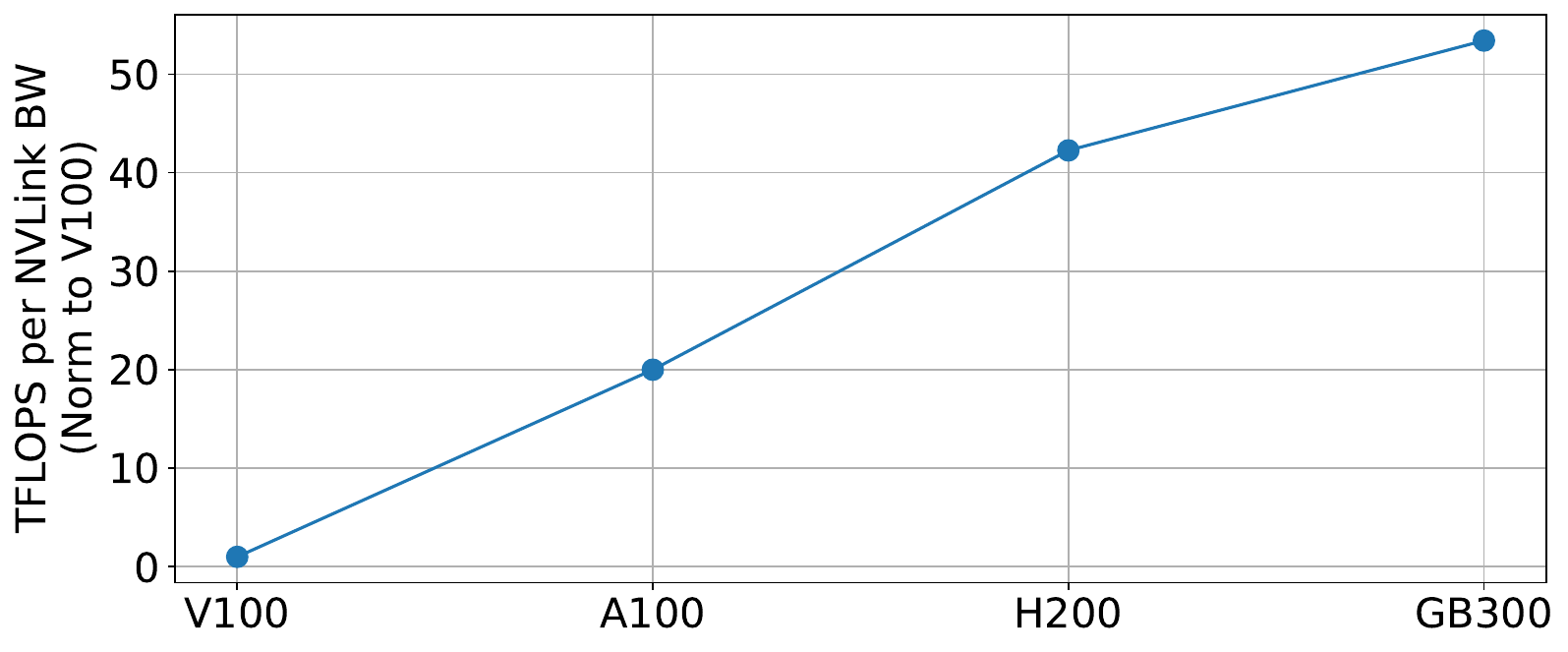}
\caption{Trends of ratio between FLOPs/second and inter-device interconnect bandwidth of recent xPUs.}
\label{fig:nvlink_flop_ratio}

\end{figure}

We define the FLOPs-per-Gbps ratio as the measure of computational performance per unit of interconnect bandwidth.
As shown in Figure~\ref{fig:nvlink_flop_ratio}, this ratio increases by approximately 2.5$\times$ from A100 to GB300, indicating that interconnect bandwidth scaling lags substantially behind computational performance improvements.

\section{Industry approach to AI Chip scaling}

Industry leaders recognize the challenges posed by limited memory capacity, memory bandwidth, and inter-device interconnect bandwidth.
Their response has been to move AI hardware toward increasingly monolithic and vertically integrated systems designed for large-scale AI workloads.
The current NVIDIA roadmap exemplifies this industry-wide shift from discrete, chip-level designs to modular, rack- and datacenter-scale architectures.
\begin{itemize}
    \item \textbf{H200 and Earlier AI Chips:}
    The NVIDIA H200 Tensor Core GPU and its predecessors follow a \emph{discrete GPU design}, with the \textbf{CPU} and \textbf{GPU} interconnected via PCIe~\cite{nvidia_h200_datasheet_2023,nvidia_h100_whitepaper_2022}.
    This modular approach allows cloud service providers to customize \textbf{CPUs}, \textbf{DRAM}, and \textbf{networking components} (e.g., InfiniBand or Ethernet) based on workload needs, maintaining flexibility across diverse AI applications.

    \item \textbf{GH200 Chip:}  
    The Grace Hopper Superchip integrates a \textbf{Grace CPU} and \textbf{Hopper GPU} using \textbf{NVLink-C2C}, enabling a unified memory system~\cite{nvidia2023gracehopper}.  
    This architecture eliminates PCIe bottlenecks and significantly boosts \emph{AI memory capacity} for large-scale AI and HPC workloads.  
    However, it reduces configurability for cloud providers, moving the design toward a tightly coupled, vendor-defined system.
    
    \item \textbf{GB200 Modular Rack:}  
    NVIDIA’s \textbf{GB200}, part of the Blackwell architecture, scales AI hardware integration to the rack level~\cite{nvidia2024blackwell}.  
    Like the GH200, two Blackwell GPUs are tightly coupled with a Grace CPU via NVLink-C2C.  
    More notably, the \textbf{GB200 NVL72} superpod interconnects up to 72 GPUs through \textbf{NVSwitch}, forming a rack-level modular system optimized for massive generative AI workloads~\cite{nvidia2024gb200nvl72}.  
    While providers retain some flexibility in networking, most of the system architecture is standardized by NVIDIA.
    
    \item \textbf{Future Datacenter-Scale Architectures:}  
    NVIDIA envisions AI-focused datacenters that integrate Grace CPUs, GPUs, memory, and networking into unified systems—transforming datacenters into \textbf{AI factories}~\cite{nvidia-ai-factory, blog-ai-factory}.  
    This approach represents a decisive move toward vertically integrated infrastructure for maximum performance and efficiency, though it would likely leave cloud providers responsible primarily for power delivery and cooling management.
\end{itemize}

\begin{figure}[h]
\centering
\includegraphics[page=9,width=0.4\linewidth]{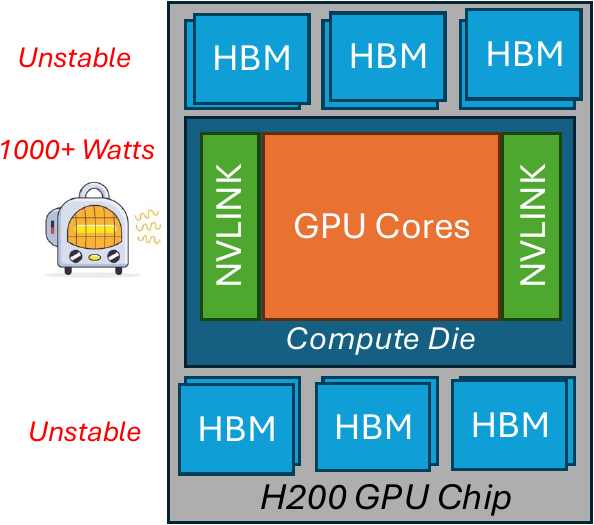}
\caption{GPU and HBM tightly coupled via a silicon interposer.}
\label{fig:gpuplushbm}
\end{figure}

A foundational design pattern in GPU hardware scaling—widely adopted across vendor roadmaps—is the tightly coupled \textbf{core–HBM–interconnect} architecture, in which compute cores and interconnects reside on the same die and connect to HBM stacks through a silicon interposer (Figure~\ref{fig:gpuplushbm}).
However, as AI workloads increasingly demand higher memory capacity and I/O bandwidth relative to compute, such designs face growing limitations.
Supporting these shifting requirements necessitates more aggressive scaling of HBM and interconnect bandwidth; yet the current packaging approach imposes three fundamental structural constraints that inhibit peripheral subsystem scaling.
\begin{itemize}
\item \textbf{Perimeter-Limited Integration:}
Increasing GPU die area by a factor of \( X \) only allows a \( \sqrt{X} \) increase in perimeter length, constraining the physical attachment area available for additional memory (HBM stacks) and interconnect I/O bandwidth.
This geometric limitation directly results in slower scaling of memory and I/O subsystems compared to compute.
\item \textbf{Thermal and Power Constraints:}
Tightly integrated systems already operate at extreme power levels (exceeding 1000~W).
Additional scaling through \emph{vertical} stacking exacerbates thermal density and power delivery challenges, leading to potential instability and limiting feasible integration density.
\item \textbf{Packaging and Reticle Limits:}
The silicon interposers used to connect HBM stacks are already approaching the physical limits of reticle and substrate sizes.
These constraints hinder further expansion of on-package memory bandwidth and capacity, particularly as model sizes continue to increase.
\end{itemize}
These scaling limitations render the traditional GPU–HBM–interconnect tightly coupled architecture increasingly unsustainable for future AI workloads.
As model trends continue to emphasize higher memory bandwidth and capacity relative to compute, overcoming these bottlenecks will require rethinking system architecture to \textbf{decouple} memory, interconnect, and compute—enabling each to scale independently and efficiently.

\section{New Memory Subsystem Design for AI}
Next-generation AI inference workloads are increasingly \emph{bottlenecked not by raw compute, but by I/O bandwidth and memory capacity} (see \cref{sec:model_vs_hardware}).
To address this imbalance, future AI infrastructure must \textbf{decouple} memory and compute resources, allowing each to scale independently.
This represents a fundamental shift from compute-centric scaling toward data-centric system design.
This decoupling concept—known as \emph{memory disaggregation}, i.e., sharing remote memory pools across processors—has been extensively studied in CPU-centric systems over the past decade.
Most existing hardware disaggregation efforts have focused on CPU memory, with technologies such as CXL-based PCIe prototypes making progress in mitigating DRAM under-utilization and resource stranding.
However, higher access latency and the difficulty of predicting memory access patterns in CPU workloads make it challenging to achieve performance comparable to traditional monolithic memory architectures.
Consequently, memory disaggregation has seen limited adoption in practical public-cloud deployments.
We argue that xPU-based AI workloads are uniquely positioned to benefit from memory disaggregation due to three converging factors:
\begin{enumerate}
\item \textbf{Hardware readiness for high-bandwidth remote memory.}
AI inference demands memory bandwidth on par with local HBM.
Modern 224 Gbps SerDes transceivers—and forthcoming 448 Gbps generations—bring disaggregated memory bandwidth increasingly close to on-package HBM performance~\cite{224gserdes}.
\item \textbf{System practicality for latency hiding.}
Unlike CPU workloads, AI workloads exhibit predictable, coarse-grained memory access patterns, well-suited for software prefetching to mask the latency of remote disaggregated memory.
\item \textbf{System advantage by collapsing communication into computation.}
A shared disaggregated memory layer across xPUs introduces a unique opportunity to merge communication with computation—an essential optimization for maximizing accelerator utilization (MFU).
By enabling direct inter-xPU data exchange through shared-memory load/store operations, the system effectively collapses communication into computation at the architectural level.
Given the predictable and block-structured data exchange patterns of AI workloads, a simple block-wise read/write memory model is sufficient to ensure both correctness and efficiency.
\end{enumerate}
Unlike traditional disaggregated systems—primarily aimed at reducing hardware stranding—\textbf{xPU memory disaggregation} directly addresses the peripheral bottlenecks constraining today’s AI infrastructure.
It provides clear system-level advantages over conventional shared-nothing, scale-up architectures.
By enabling scalable, decoupled memory pools and collapsing inter-xPU communication into shared-memory operations, this approach delivers \emph{greater flexibility}, \emph{resilience against rapidly evolving workloads}, and \emph{superior utilization} of GPU accelerators.
As AI models continue to grow in complexity and scale, memory disaggregation will play a pivotal role in overcoming systemic bottlenecks—unlocking efficiency and scalability that legacy architectures cannot achieve.
\chapter{\phoenix System Design}
\section{Overview -- A Shared-memory AI-chip Scaling Architecture}
\label{sec:arch}

\begin{figure}[h]
\begin{minipage}[b]{0.4\linewidth}
\centering
\includegraphics[page=10,width=0.95\linewidth]{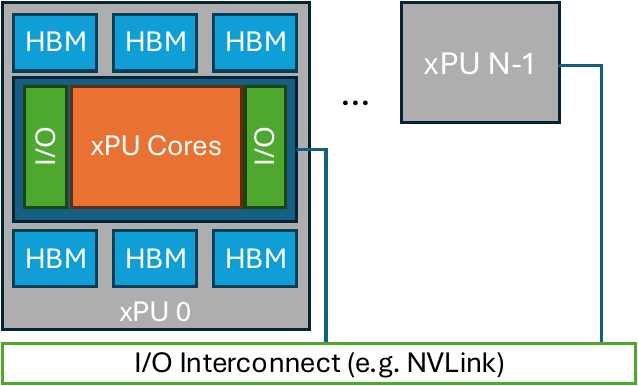}
\end{minipage}
\hfill
\begin{minipage}[b]{0.6\linewidth}
\centering
\includegraphics[page=11,width=0.95\linewidth]{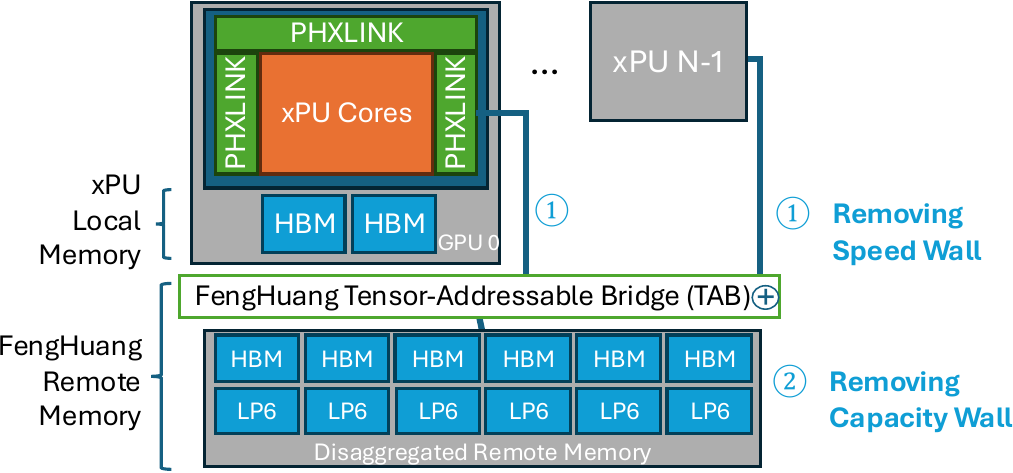}
\end{minipage}

\vspace{-0.5em} 
\begin{minipage}[t]{0.4\linewidth}
\centering
\captionof{figure}{Conventional Shared-nothing Scale-up Architecture}
\label{fig:shared-nothing}
\end{minipage}
\hfill
\begin{minipage}[t]{0.6\linewidth}
\centering
\captionof{figure}{\phoenix Shared-memory Scale-up Architecture}
\label{fig:shared-mem}
\end{minipage}
\end{figure}

\phoenix introduces a shared-memory architecture that departs from the traditional scale-out model of shared-nothing AI accelerators (xPUs). 
Unlike conventional designs—built around proprietary communication interconnects (e.g., NVLink) and heavy software orchestration (e.g., MPI) (Figure~\ref{fig:shared-nothing})—\phoenix allows xPUs to access both local and remote memory through a unified memory fabric, i.e., \tabfull (\tab). 
This approach eliminates long-standing peripheral walls that have constrained xPU scalability (Figure~\ref{fig:shared-mem}).

In this architecture, each xPU retains its own local high-bandwidth memory while seamlessly accessing a shared pool of \remotemem through the \tab fabric. 
This decoupled design supports both capacity scaling and inter-xPU communication using standard memory semantics, delivering improved scaling efficiency for modern AI workloads.

In the current \phoenix configuration, we distinguish two types of memory areas (Figure~\ref{fig:shared-mem}):

\begin{itemize}
\item \textbf{\localmem:} Certain HBM units are retained locally within each xPU, serving as a cache for \phoenix remote memory where frequently accessed tensors reside. From the kernel software’s perspective, local memory behaves like existing HBM, but its capacity and bandwidth are tuned to workload characteristics for efficient caching and computation.

\item \textbf{\remotemem:} Each xPU’s memory capacity is extended through the \tab, forming the so-called \remotemem. Access to this remote memory resembles peer-to-peer GPU memory access over NVLink. Tensors in remote memory can be copied into local memory without requiring cache coherence, or accessed by the SMs through the caching hierarchy. Moreover, \remotemem serves as a centralized memory pool for efficient aggregation of inter-chip communication traffic\footnote{The aggregation operation is physically performed by the \tab, enabling \phoenix hardware to use standard memory modules directly.}. We will later show that this shared-memory architecture accelerates inter-GPU communication by up to two orders of magnitude.
\end{itemize}

\phoenix fundamentally redesigns the xPU memory subsystem by incorporating a shared pool of high-speed remote memory accessible through a unified memory fabric. This fabric enables xPUs to access remote memory with \emph{comparable} bandwidth to their local HBM. \phoenix represents a significant departure from traditional shared-nothing scale-up architectures, addressing key peripheral bottlenecks (\cref{chap:background}) of xPU systems as follows:

\begin{itemize}
\item \textbf{Memory Capacity Wall:} \phoenix decouples compute scaling from memory capacity through the \tabfull. The northern and southern sides of the \tab can scale independently according to workload demands. Specifically, the \remotemem tier alleviates the increasing memory pressure from modern AI workloads, as discussed in \cref{sec:computation_hbm_cap}.

\item \textbf{Memory Bandwidth Wall:} A core advantage of \phoenix is its separation of memory bandwidth and capacity into distinct tiers. The \localmem tier is optimized for bandwidth, storing tensors required for immediate execution, while the \remotemem tier is optimized for capacity, storing parameters and intermediate tensors not immediately needed. This architectural decoupling enables more efficient physical design of \localmem, addressing the bandwidth bottleneck prevalent in LLM inference workloads (\cref{sec:memory_bandwidth_wall}).

\item \textbf{Communication Bandwidth Wall:} Traditional xPU scale-up architectures aggregate HBM across devices via interconnects typically an order of magnitude slower than local HBM. \phoenix mitigates this limitation by using \remotemem as the communication medium, which offers bandwidth comparable to local HBM. Furthermore, \phoenix enables direct memory read and write operations across xPUs on shared memory regions, greatly reducing communication overhead compared with MPI-based protocols used in shared-nothing systems.
\end{itemize}

In the next two sections, we examine two key hardware design features in \phoenix: (1) mechanisms to hide the long memory access latency to \remotemem, and (2) a redesigned inter-xPU communication model based on shared memory.

\section{Hardware Feature 1: Tensor Prefetcher for Local–Remote Memory Paging}
\label{sec:tensor_prefetcher}

\phoenix introduces a hierarchical memory design composed of both local and remote memory tiers, forming a \emph{two-tier memory architecture}. 
This structure enables independent scaling of memory bandwidth and capacity, tailored to the needs of modern AI inference workloads. 
However, disaggregating memory inevitably increases access latency. 
Directly feeding data from \remotemem into xPU cores—without additional hardware or software support-could result in significant performance degradation.

To mitigate this latency, \phoenix incorporates a dedicated tensor prefetching mechanism to overlap remote memory access with computation. 
While traditional CPU workloads present challenges for prefetching due to unpredictable access patterns, AI inference workloads are significantly more regular; their execution graphs are largely static and operator sequences are known in advance. 
This predictability makes prefetching not only feasible but highly effective.

To exploit this property, \phoenix integrates a \prefetcher, a critical component designed to hide disaggregated memory access latency through proactive management of tensor movement between remote and local memory.

\begin{figure}[h]
\centering
\includegraphics[page=4,width=1.0\linewidth]{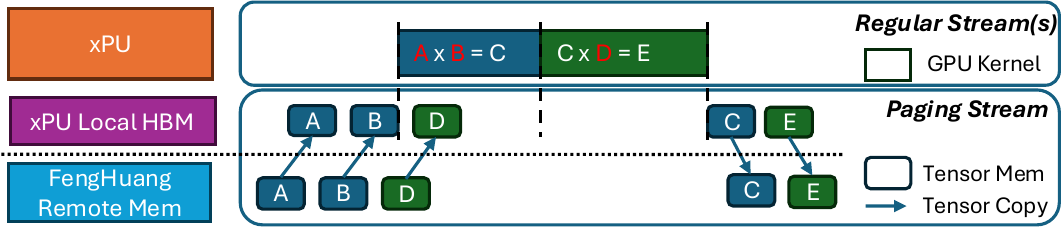}
\caption{\prefetcher -- \localmem as the paging memory of the \remotemem}
\label{fig:sim-method}
\end{figure}

As shown in Figure~\ref{fig:sim-method}, the \phoenix system consists of three main components: the xPU compute cores, the \localmem, and the \remotemem. 
The compute cores and \localmem operate similarly to those in conventional GPU architectures. 
The novelty lies in how memory is managed across the local–remote boundary.

\phoenix defines two execution streams:
\begin{itemize}
    \item \textbf{Regular Stream(s):} These execute xPU kernels using the existing xPU software stack.
    \item \textbf{Paging Stream:} A dedicated background stream that orchestrates data movement by prefetching tensors from \remotemem into \localmem ahead of their use by the Regular Stream(s) and evicting tensors that are no longer needed.
\end{itemize}

The \prefetcher coordinates the Paging Stream to ensure that each kernel’s working set is available in \localmem ahead of execution. 
When local memory space becomes constrained, tensors with low reuse likelihood are paged out to \remotemem. 
In the ideal case, the Paging Stream runs fully in parallel with computation, effectively hiding \remotemem latency and enabling high utilization of both compute and memory bandwidth resources.

\section{Hardware Feature 2: \phoenix Shared Memory for Inter–xPU Communication}
\label{sec:shared_remote_mem}

In addition to the xPU’s memory expansion, \remotemem is designed to enable high-performance communication between xPUs through \emph{shared} remote memory and the \tab (\cref{fig:shared_memory_design}). 
This approach eliminates the dependency on proprietary GPU interconnects such as NVLink, enabling a scalable and vendor-agnostic architecture.

\subsection{\phoenix Shared Memory Operations}
\begin{figure}[h]
    \centering
    \includegraphics[page=13, width=0.7\textwidth]{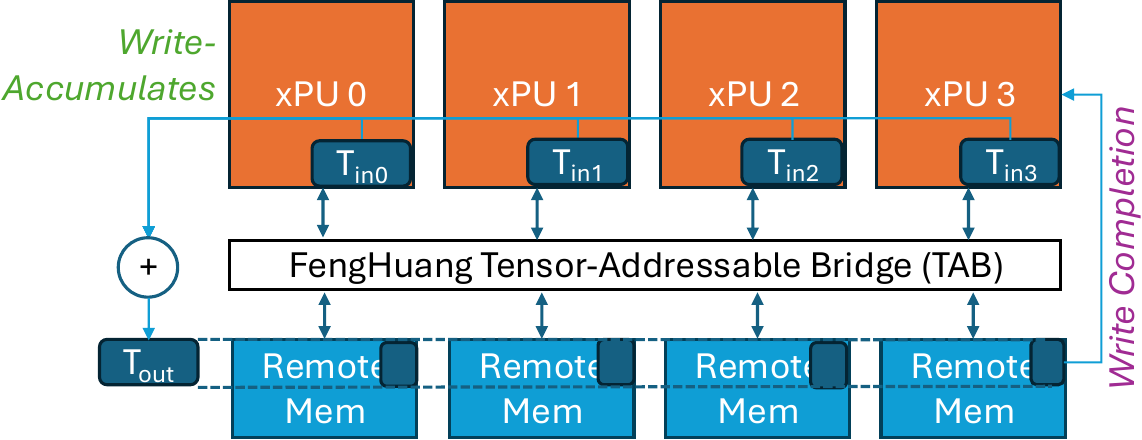}
    \caption{\phoenix \remotemem Design for Inter-xPU Communication. }
    \label{fig:shared_memory_design}
\end{figure}

In addition to standard reads and writes, \phoenix introduces two novel memory operations to enhance inter–xPU communication efficiency for AI workloads (\cref{fig:shared_memory_design}):
\begin{itemize}
    \item \textbf{Write-Accumulate:} Enables direct accumulation of tensor data in \remotemem, which is shared across all xPUs connected to the \tab. 
    Each write-accumulate operation is associated with an address and a data value; the incoming data is accumulated with the existing data residing in shared memory. 
    The write-accumulate operation is designed for reduction workloads, where each step is commutative. This property simplifies the hardware design, as it does not require strict write ordering.
    \item \textbf{Write Completion Notification:} Ensures synchronization among xPUs by signaling when data writes have completed.
\end{itemize}

The system also employs a uniform data layout, evenly striping tensors across all memory modules to maximize bandwidth utilization.

\subsection{Implementation of the Five Communication Operations on \phoenix}
The \remotemem architecture supports five fundamental communication operations: \textbf{AllReduce}, \textbf{ReduceScatter}, \textbf{AllGather}, \textbf{AllToAll}, and \textbf{P2P Send/Recv}. 
These operations are essential for efficient communication in distributed AI workloads such as deep learning training and inference. 
Below, we illustrate how \phoenix implements these operations using its \remotemem architecture.

\begin{figure}[h]
    \centering
    \includegraphics[page=15, width=0.95\textwidth]{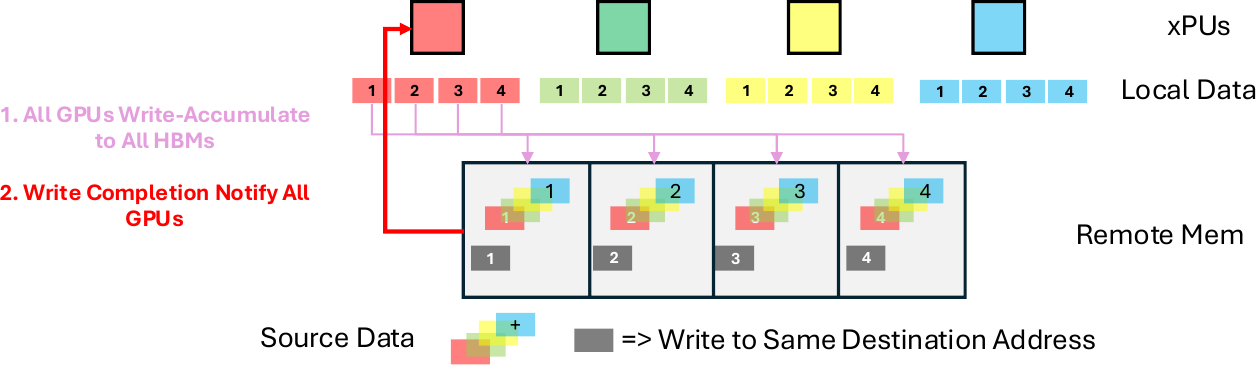}
    \caption{AllReduce and ReduceScatter operations on \phoenix}
    \label{fig:allreduce_reducescatter}
\end{figure}

\subsubsection{AllReduce and ReduceScatter (Figure~\ref{fig:allreduce_reducescatter})}
AllReduce and ReduceScatter leverage the \tabfull to perform accumulation and scattering efficiently. Their execution proceeds as follows:
\begin{enumerate}
    \item Each xPU partitions its local data into multiple chunks.
    \item All xPUs perform write-accumulate operations on their corresponding chunks to the shared memory banks in parallel.
    \item After all write-accumulate operations are complete, the \tab sends a completion signal to synchronize all participating xPUs.
\end{enumerate}

AllReduce and ReduceScatter differ only in how their result tensors are consumed. 
In AllReduce, all xPUs read the same aggregated tensor, whereas in ReduceScatter, each xPU reads its corresponding portion of the result tensor.

\begin{figure}[h]
    \centering
    \includegraphics[page=16, width=0.95\textwidth]{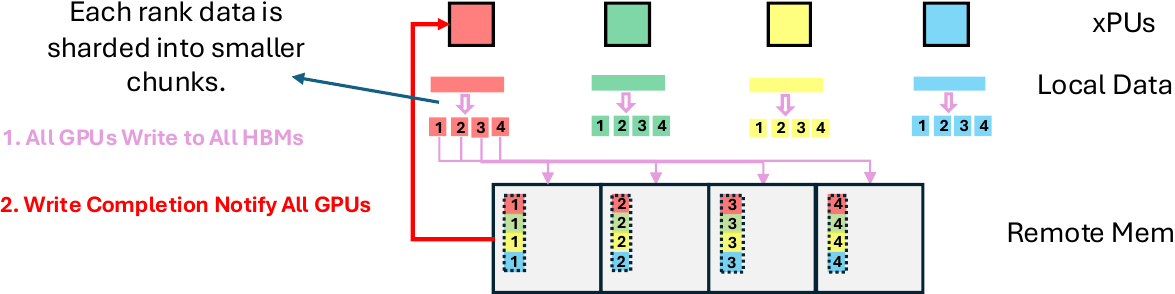}
    \caption{AllGather and AllToAll operations on \phoenix }
    \label{fig:allgather_alltoall}
\end{figure}

\subsubsection{AllGather and AllToAll (Figure~\ref{fig:allgather_alltoall})}
\phoenix implements AllGather and AllToAll by distributing and retrieving data efficiently across xPUs:
\begin{enumerate}
    \item Each xPU shards its local data into multiple smaller chunks.
    \item After all xPUs complete writing, the \tab sends a write-completion notification for synchronization.
\end{enumerate}

AllGather and AlltoAll only differs in how result tensors are consumed later. For AllGather, xPUs read the same result tensor. For AlltoAll, xPUs read their respective result tensor chunks.

\begin{figure}[h]
    \centering
    \includegraphics[page=17, width=0.4\textwidth]{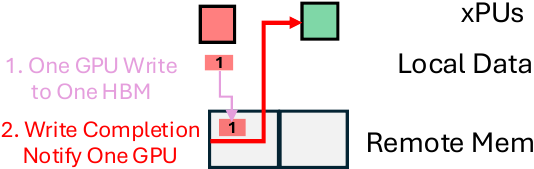}
    \caption{P2P Send/Recv operations on \phoenix}
    \label{fig:p2p_send_recv}
\end{figure}

\subsubsection{P2P Send/Recv (Figure~\ref{fig:p2p_send_recv})}
Point-to-point (P2P) communication in \phoenix leverages \remotemem for scalable GPU-to-GPU communication:
\begin{enumerate}
    \item The sending GPU writes its local data to a designated shared memory location.
    \item The \tab sends a write-completion notification to the receiving xPU once the data is available, ensuring synchronization.
\end{enumerate}

\subsection{Theoretical Analysis of \phoenix Speed-up over NVLink }

The performance improvements of \phoenix Shared Memory are driven by two key enablers: \textbf{reduced data movement} and \textbf{superior link performance}. This section analyzes these enablers in both \textbf{latency-bound} and \textbf{bandwidth-bound} scenarios, followed by an overall summary table.

\subsubsection{Analysis Setup}

In \phoenix, we assume the following minimal operation latency as follows in \cref{tab:latency_breakdown}.

\begin{table}[h]
    \centering
    \begin{tabular}{|l|c|c|}
        \hline
        \textbf{Operation} & \textbf{Latency Component} & \textbf{Time (ns)} \\
        \hline \hline
        \multirow{6}{*}{Read} 
        & Read command from GPU to \phoenix & 40 \\
        & Read command processing in \phoenix & 10 \\
        & Read command from \phoenix to remote HBM & 40 \\
        & Remote HBM read time & 50 \\
        & Data from remote HBM to \phoenix & 40 \\
        & Data from \phoenix to GPU & 40 \\
        \hline
        \textbf{Total Read Latency} & & \textbf{220} \\
        \hline \hline
        \multirow{3}{*}{Write (Post Write Scheme)} 
        & Write command and data from GPU to \phoenix & 40 \\
        & Write command processing in \phoenix & 10 \\
        & Write completion notification from \phoenix to GPU & 40 \\
        \hline
        \textbf{Total Write Latency} & & \textbf{90} \\
        \hline \hline
        \textbf{Atomic operation completion notification} & & \textbf{40} \\
        \hline 
    \end{tabular}
    \caption{Breakdown of Minimal Operation Latency (2KB data) in \phoenix System.}
    \label{tab:latency_breakdown}
\end{table}

The total latency for read, write,  write-accumulate, and write-completion operations in the \phoenix system can be expressed using the following equations. In particular, the $\text{data\_size}$ is the size of data transfer in \verb|bytes|, and the $\text{bandwidth}$ is the \phoenix memory crossbar transfer rate between a GPU and the shared memory in \verb|GBytes/s|.

\begin{equation}
    \text{Read Latency} = 220\,\text{ns} + \frac{\text{data\_size}}{\text{bandwidth}}
    \label{eq:read_latency}
\end{equation}

\begin{equation}
    \text{Write Latency} = 90\,\text{ns} + \frac{\text{data\_size}}{\text{bandwidth}}
    \label{eq:write_latency}
\end{equation}

\begin{equation}
    \text{Write Accumulate Latency} = 90\,\text{ns} + \frac{\text{data\_size}}{\text{bandwidth}}
    \label{eq:write_accumulate_latency}
\end{equation}

\begin{equation}
    \text{Write-Completion Notification Latency } = 40\,\text{ns}
    \label{eq:optimized_write_accumulate_latency}
\end{equation}

The overall \phoenix setup (Figure~\ref{fig:phoenix_simulation_setup} assumes a system of GPUs connected via the \phoenix Controller with shared HBM memory for tensor communication. Key configurations include:

\begin{figure}[h]
    \centering
    \includegraphics[page=18, width=0.95\textwidth]{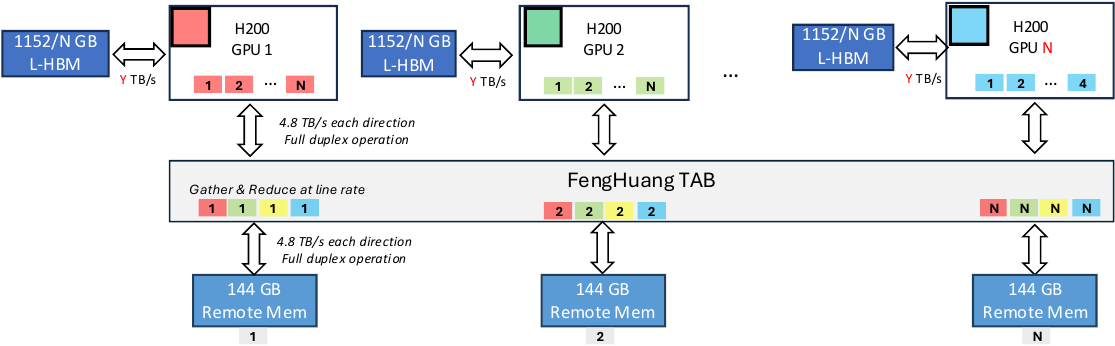}
    \caption{\phoenix Shared-Memory Crossbar Simulation Setup}
    \label{fig:phoenix_simulation_setup}
\end{figure}

\begin{itemize}
    \item \textbf{Baseline System:}
    \begin{itemize}
        \item \textbf{Compute:} 8 H200 GPUs.
        \item \textbf{Memory:} Total of 1152 GB of HBM operating at a bandwidth of 38.4 TB/s.
        \item \textbf{Communication:} NVLink interconnect with a bi-directional bandwidth of 900 GB/s per GPU.
    \end{itemize}
    \item \textbf{\phoenix System:}
    \begin{itemize}
        \item \textbf{Compute:} $N$ H200 GPUs.
        \item \textbf{Memory:} Total of $144 \times N$ GB of HBM shared across GPUs, scaled at $Y \times N$ TB/s based on GPU count.
        \item \textbf{Communication:} A 4.8 TB/s bi-directional crossbar per GPU, integrated with in-memory tensor reduction through the \phoenix Controller.
    \end{itemize}
\end{itemize}

The following lists how we set up parameter for \phoenix.
\begin{itemize}
\item We choose \textbf{N} in \phoenix to be \textbf{8}, the same as the baseline system, because we only evaluate the communication subsystem of \phoenix in comparison to be the baseline.

\item We choose the \textbf{bandwidth} in \cref{eq:read_latency}, \cref{eq:write_latency} and \cref{eq:write_accumulate_latency} to be \textbf{4.0 TB\/s}, factoring in the typical hardware efficiency in communication system.

\item We \emph{ignored} the setting of \textbf{Y}, the local HBM bandwidth, since we only consider inter-GPU communication in our evaluation, which uses \emph{direct remote memory access}.
\end{itemize}

\subsubsection{Performance Enabler 1: Reduced Data Movement}

\phoenix minimizes data movement by using in-memory tensor reduction, drastically reducing total data transfer.

\begin{itemize}
    \item \textbf{Latency-Bound Scenarios (Small Tensor Sizes)}

In latency-bound scenarios, small tensors dominate, and the key metric is the number of transfers required:
\[
\text{\# of Data Transfers (NVLink)} = 2(N-1) \quad \text{and} \quad \text{\# of Data Transfer (\phoenix)} = 1
\]
\[
\text{Speed-Up}_{\text{Enabler 1 (Latency-Bound)}} = \frac{\text{Data Transfer (NVLink)}}{\text{Data Transfer (\phoenix)}} = \frac{2(N-1)}{1}
\]
For \( N = 8 \), this becomes:
\[
\text{Speed-Up}_{\text{Enabler 1 (Latency-Bound)}} = 2(8-1) = 14\times
\]

\item \textbf{Bandwidth-Bound Scenarios (Large Tensor Sizes)}

In bandwidth-bound scenarios, large tensors dominate, and the transfer bandwidth is critical:
\[
\text{Data Transfer (NVLink)} = 2(N-1) \times T/N \quad \text{and} \quad \text{Data Transfer (\phoenix)} = T~\footnote{Data transfer is calculated per GPU, and the NVLink baseline uses ring-allreduce algorithm.}
\]
\[
\text{Speed-Up}_{\text{Enabler 1 (Bandwidth-Bound)}} = \frac{\text{Data Transfer (NVLink)}}{\text{Data Transfer (\phoenix)}} = \frac{2(N-1) \times T/N}{T}
\]
For \( N = 8 \), this becomes:
\[
\text{Speed-Up}_{\text{Enabler 1 (Bandwidth-Bound)}} = \frac{2(8-1) \times T/8}{T} = 14 / 8 = 1.75\times
\]
\end{itemize}

\subsubsection{Performance Enabler 2: Superior Link Performance}

\phoenix achieves superior link performance through its high-bandwidth crossbar interconnect, which reduces latency and increases bandwidth.

\begin{itemize}

\item \textbf{Latency-Bound Scenarios (Small Tensor Sizes)}

In latency-bound scenarios, the performance is limited by fixed latencies per transfer:
\[
\text{Latency (NVLink)} = 1000 \, \text{ns (read)} \quad \text{and} \quad \text{Latency (\phoenix)} = 220 \, \text{ns (read)}
\]
\[
\text{Latency (NVLink)} = 500 \, \text{ns (write)} \quad \text{and} \quad \text{Latency (\phoenix)} = 90 \, \text{ns (write)}
\]
\[
\text{Speed-Up}_{\text{Enabler 2 (Latency-Bound)}} = \frac{\text{Latency (NVLink)}}{\text{Latency (\phoenix)}} = \frac{1000}{220} or \frac{500}{90} \approx 5\times
\]

\item \textbf{Bandwidth-Bound Scenarios (Large Tensor Sizes)}

In bandwidth-bound scenarios, the performance is determined by the transfer bandwidth:
\[
\text{Bandwidth (NVLink)} = 450 \, \text{GBps} \quad \text{and} \quad \text{Bandwidth (\phoenix)} = 4800 \, \text{GBps}
\]
\[
\text{Speed-Up}_{\text{Enabler 2 (Bandwidth-Bound)}} = \frac{\text{Bandwidth (\phoenix)}}{\text{Bandwidth (NVLink)}} = \frac{4000}{450} = 8.89\times
\]

\end{itemize}

\subsubsection{Overall Speed-Up}
The overall speed-up of \phoenix is the product of the speed-ups from Enabler 1 and Enabler 2:
\begin{itemize}

\item \textbf{Latency-Bound Scenarios:}

\[
\text{Overall Speed-Up}_{\text{Latency-Bound}} = 14 \times 5 = 70\times
\]

\item \textbf{Bandwidth-Bound Scenarios:}
\[
\text{Overall Speed-Up}_{\text{Bandwidth-Bound}} \approx 1.75 \times 8.89 = 15.56\times
\]
\end{itemize}

\section{High-Level Software Design}

The \phoenix software stack is designed to seamlessly port existing AI applications onto the \phoenix architecture with small, manageable code changes. 
Since the \phoenix architecture reuses much of the existing xPU chip design, applications can similarly reuse most of the existing xPU software stack for computational operations that operate over \localmem (e.g., matrix multiplication).\footnote{Adopting \phoenix may change existing xPU chip configurations—for example, increasing local memory bandwidth or core counts. Such changes may require further tuning of low-level implementations to fully exploit the hardware. However, these considerations are orthogonal to the \phoenix design and are not discussed in this paper.} 
The key role of the \phoenix software stack is to \emph{page in} and \emph{page out} tensors between \localmem and \remotemem in the background while the xPU executes local computations. 
The software stack also provides efficient communication primitives that operate over the \tab, replacing existing xPU interconnects such as NVLink.

The \phoenix software stack provides multiple interface layers, allowing developers to choose the abstraction level that best suits their needs. 
In the remainder of this section, we use the GPU software stack as an example, since GPUs are the most representative xPU for AI workloads today. 
The \phoenix software stack consists of the following components:

\begin{itemize}
    \item \textbf{Extended GPU Instruction Set Architecture (ISA).} 
    The \phoenix ISA extends the existing GPU ISA, enabling GPU threads to (1) directly access \remotemem, (2) initiate the hardware engine to copy data between \localmem and \remotemem, and (3) invoke the hardware accumulator on the \tab. 
    These instructions are called from GPU kernel code, enabling optimization of data movement during computation.
    
    \item \textbf{GPU Kernel Library.} 
    The \phoenix GPU kernel library is a header-only collection of reusable GPU kernel implementations built on the \phoenix ISA. 
    It provides kernel-level interfaces (e.g., CUDA device functions) for copying data between \localmem and \remotemem, as well as collective communication operations such as AllReduce, ReduceScatter, AllGather, AllToAll, and P2P Send/Recv.
    
    \item \textbf{Runtime API.} 
    The \phoenix runtime API offers a host-side interface for managing remote memory—including allocation, deallocation, and synchronization—and provides host-side wrappers for the \phoenix GPU kernel library, allowing developers to invoke GPU kernels from host code.
    
    \item \textbf{PyTorch Extension.} 
    The \phoenix PyTorch extension implements \textit{pageable tensor} data structures and operations using the \phoenix runtime API. 
    It provides a Python interface for paging tensors in and out of memory, enabling seamless integration with existing PyTorch code. 
    The extension is fully compatible with popular PyTorch-based inference frameworks such as vLLM and SGLang. 
    Using this extension, we also implement automatic paging and prefetching mechanisms, allowing existing PyTorch code to run on a \phoenix platform with minimal code changes—users simply apply a provided Python decorator to the PyTorch modules or functions they wish to make pageable.
\end{itemize}

\chapter{Evaluation}
In this section, we evaluate the performance of representative workloads on the \phoenix-based GPU architecture. 
Our objective is to analyze how different configurations of \phoenix’s local–remote memory hierarchy impact overall workload efficiency.

\section{Setup}

\subsection{Hardware Setup}

\noindent\textbf{\phoenix.} 
\phoenix consists of four H200 GPUs, each providing a 1.33× compute improvement and a 1.5×–2.0× local memory speedup (see \cref{tab:xpu_spec}). 
These GPUs are interconnected through the \phoenix TAB, offering a total remote memory capacity of 1,152~GB (see \cref{tab:scale_up_spec}).

\noindent\textbf{Baseline.} 
The baseline configuration consists of a single Nvidia H200 node with eight GPUs interconnected via an NVLink~4.0 switch. 
The total memory capacity is matched to \phoenix at 1,152~GB for a fair comparison.

\begin{table}[htbp]
\centering
\small
\begin{tabular}{ccccc}
\toprule
\textbf{Systems} & \textbf{Total xPUs} & \textbf{Compute} & \textbf{Local HBM Speed} & \textbf{Local HBM Capacity} \\ \midrule
FH4-1.5xM & 4 & 1.33$\times$H200 & 7.2TB/s (1.5x H200) & As much as needed \\
FH4-2.0xM  & 4 & 1.33$\times$H200 & 9.6TB/s (2.0x H200) & As much as needed \\
Baseline8     & 8 & H200             & 4.8TB/s              & 144GB \\
\bottomrule
\end{tabular}
\caption{System and xPU Specifications}
\label{tab:xpu_spec}
\end{table}

\begin{table}[htbp]
\centering
\small
\begin{tabular}{cccccc}
\toprule
\multirow{2}{*}{\textbf{Systems}} & \multirow{2}{*}{\textbf{Scale}} & \multirow{2}{*}{\textbf{Hardware}} & \textbf{Bandwidth} & \multirow{2}{*}{\textbf{Latency}} & \textbf{Remote Memory} \\ 
                 &                &                   & (per GPU)           &       & \textbf{Capacity} \\ \midrule
FH4-1.5xM & 4 & \phoenix TAB & 4.0--6.4TB/s & Same as NVLink 4.0 & 1152 GB \\
FH4-2.0xM  & 4 & \phoenix TAB & 4.0--6.4TB/s & Same as NVLink 4.0 & 1152 GB \\
Baseline8     & 8 & NVLink 4.0  & 450GB/s       & $\sim$1000ns Read / $\sim$500ns Write~\tablefootnote{\emph{Measured in real systems}} & -- \\ 
\bottomrule
\end{tabular}
\caption{Scale-up Network Specifications}
\label{tab:scale_up_spec}
\end{table}

\subsection{Software Setup}

\noindent\textbf{Workloads.} 
We evaluate three inference workloads—GPT-3 175B, Grok-1, and Qwen-3 235B—which represent three typical large language model (LLM) architectures. 
GPT-3 175B represents a traditional dense Transformer architecture. 
Grok-1 is a sparse Mixture-of-Experts (MoE) model in which each expert is a replica of the original Transformer feed-forward network (FFN). 
Each layer contains eight experts with top-2 expert activation. 
Qwen-3 235B follows the approach of DeepSeek, adopting a large-scale, fine-grained expert architecture in which each expert’s intermediate size is significantly smaller than the original FFN. 
Each layer includes 128 experts, and eight of them are activated for each token.

\noindent\textbf{Inference Tasks.} 
We evaluate two categories of inference tasks: traditional question-answering (Q\&A) and reasoning. 
The primary difference lies in the relationship between prompt length and generation length. 
For end-to-end evaluation, we use fixed (prompt, generation) length pairs of (4,096, 1,024) tokens for Q\&A and (512, 16,384) tokens for reasoning, with a batch size of eight.

\noindent\textbf{Inference Framework.} 
We use the SGLang framework (v0.4.10.post2). 
Key configurations include FlashAttention-3 for attention and a Triton-based FusedMoE implementation for MoE layers. 
As \phoenix focuses on the GPU-side memory architecture, we exclude non-GPU-side optimizations such as request scheduling from our evaluation.

\noindent\textbf{Metrics.} 
Latency is used as the primary performance metric. 
We report both the time to first token (TTFT) and the time per output token (TPOT), comparing \phoenix with the baseline configuration.

\subsection{Simulation Methodology}

We developed a simulator to model \phoenix’s behavior, using Nsight profiling data from baseline GPU runs. 
The simulator constructs a dependency graph from profiling traces and inserts prefetching nodes. 
We adopt a simple lookahead-1 prefetching strategy, which overlaps data transfer with computation. 
Each node in the dependency graph initiates prefetching for its immediate successor in the execution order, maintaining a prefetch window of \(w = 1\). 
This approach also reduces pressure on local memory capacity, since only the minimum required data are stored locally. 

To more accurately model prefetching overhead, we apply a scaling coefficient to the theoretical remote memory bandwidth, similar to empirical NVLink behavior. 
In particular, larger tensor sizes achieve higher effective bandwidth and exhibit reduced latency dominance.

\begin{equation}
    \text{Prefetching Overhead} = \frac{\text{Tensor Size}}{\text{Remote Memory Bandwidth} \times \text{Efficiency(Tensor Size)}}
\end{equation}
\begin{figure}[h]
 \centering
    \includegraphics[width=1.0\linewidth]{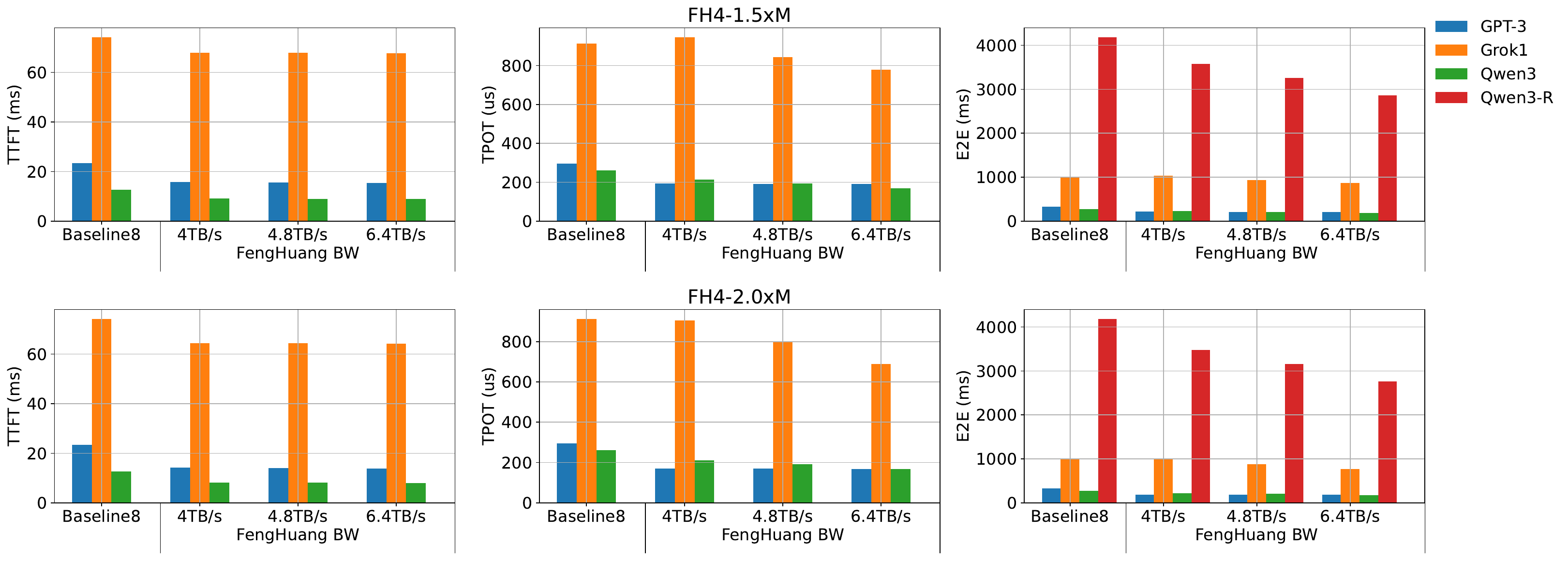}
    \caption{FengHuang vs Baseline8. Prompt length is 4096. For E2E results, we use (4096, 1024) for GPT-3, Grok1 and Qwen3 and (512, 16384) for Qwen3-R.}
    \label{fig:ttft_tpot_e2e}
\end{figure}

\section{Workload Performance}

Figure~\ref{fig:ttft_tpot_e2e} presents the TTFT, TPOT, and end-to-end latency results for Baseline and \phoenix across various remote memory bandwidth configurations.

\noindent\textbf{TTFT.} 
For a prompt length of 4,096 tokens and a remote memory bandwidth of 4.0~TB/s, FH4-1.5×M outperforms Baseline by 32.5\%, 8.4\%, and 28.9\% for GPT-3, Grok-1, and Qwen-3, respectively. 
Notably, TTFT remains relatively stable as remote memory bandwidth increases from 4.0~TB/s to 6.4~TB/s. 
This behavior arises because the prefill phase is computation-intensive, providing sufficient opportunity for prefetching to overlap completely with computation, thus incurring negligible overhead.

\noindent\textbf{TPOT.} 
The decode phase imposes higher demand on remote memory bandwidth due to its lower computational intensity, which reduces opportunities for effective prefetching overlap with computation. 
Consequently, as remote memory bandwidth increases from 4.0~TB/s to 6.4~TB/s, TPOT improvements become more pronounced, with average reductions ranging from 16\% to 28\%. 
Among the three models, Grok-1 experiences a slight slowdown at 4.0~TB/s remote memory bandwidth, primarily due to its large expert architecture. 
Furthermore, because the decode phase is inherently memory bandwidth-bound, improvements in local memory bandwidth also yield substantial reductions in TPOT.

\noindent\textbf{E2E Latency.} 
For traditional Q\&A workloads, all three models achieve performance comparable to the Baseline configuration once remote memory bandwidth reaches 4.8~TB/s. 
For the decoding-dominant Qwen-3-R workload, significant performance improvements are already observed at 4.0~TB/s remote memory bandwidth.

\noindent\textbf{Local Memory Capacity.} 
Under our prefetching strategy, the required local memory capacity is summarized in Table~\ref{tab:local_mem_cap_req}. 
This capacity requirement is determined by the peak memory usage observed during execution on the \phoenix system.

\begin{table}[h]
\centering
\begin{tabular}{@{}lll@{}}
\toprule
Model & Memory Requirement (GB) \\ \midrule
GPT-3 & 10 \\
Grok1 & 18 \\
Qwen3 & 20 \\
Qwen3-R & 20 \\
\bottomrule
\end{tabular}
\caption{Local memory capacity requirement of each workload in different phase.}
\label{tab:local_mem_cap_req}
\end{table}

\chapter{Chip-level Physical Design Framework}

A lot of \phoenix’s achievements are rooted in our rethinking of the modern xPU architectures and the novel TAB chip.  \phoenix system design enables a low-cost framework that scales from low complexity to state-of-the-art language model implementations. It increases xPU utilization to reduce xPU count, reduces local HBM to minimize xPU and HBM system in package complexity, enables remote memory at high bandwidth without the need for state-of-the-art packaging paradigms, provides the ability to incorporate LPDDR5 memory as a low-cost alternative to HBM. Many of these cost reduction vectors are being explored in the AI area~\cite{google2025hotchip, intel2025aichip}, so the \phoenix system design accelerates this momentum for the most complex of LLM workloads at warehouse scale AI computing. 

The primary focus of the \phoenix framework is to minimize the I/O bandwidth constraint associated with the local HBM while minimizing cost.  As per publicly available HBM roadmap projections, the best case scenario for Classical Inferencing Frameworks is to achieve 500GB local HBM capacity with 8 cubes coupled to 2x GPU in a single silicon interposer with an aggregate bandwidth of 50TB/sec in 2029-2030 timeframe~\cite{kaist2025hbmroadmap}. This is a 100:1 ratio of bandwidth to capacity measured as TB/sec per TB. \phoenix two tier memory orchestration requires 20GB of  local memory at a bandwidth of 10 TB/sec. This is a ratio of 500 TB/sec per TB – a 5x increase over the Classical Inferencing Frameworks. Next generation architectures, one of which can be a vertically connected DRAM to GPU, or others will come into existence in this time frame which will bring the bandwidth from 100 TB/sec per TB to 500 TB/sec/per TB~\cite{nvidia2025davies, gatech20253ddram} with lower capacity over-provisioning. \phoenix builds on this expected development to develop a new concept of AI inferencing infrastructure for the 2028-2030 deployment timeframe.

\begin{figure}[h]
\centering
\includegraphics[width=0.9\linewidth]{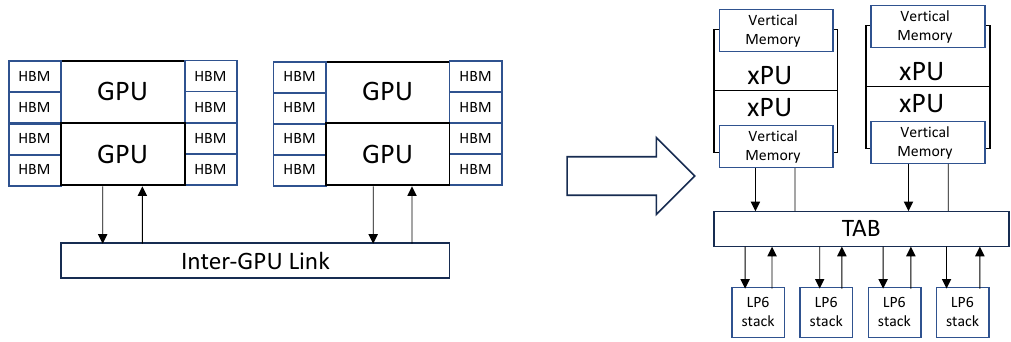}
\caption{Transition from classical inferencing framework to \phoenix with vertically connected  GPU DRAM memories in 2028-2029 timeframe~\cite{gatech20253ddram}.}
\label{fig:physical}
\end{figure}

The second focus of the \phoenix framework is the memory capacity expansion via two-tier memory orchestration. The second tier of memory is connected via the Tensor Acceleration Bridge TAB). It decouples GPU compute from memory capacity. There are many possibilities for TAB connected memory, but a promising approach is LPDDR6 stacks where all the dies are activated in parallel. TAB memory capacity can be as high as of 4096TB, an 8$\times$ increase over 500GB HBM in the public roadmaps. Each GPU is connected to the TAB via 11.5 TB/sec – 23TB/sec power and latency optimized serdes interconnect, a 4$\times$-8$\times$ improvement over the public roadmaps~\cite{kaist2025hbmroadmap}. The \phoenix two-tier memory orchestration minimizes the I/O bottlenecks associated with HBM capacity, HBM bandwidth, and NVLINK bandwidth for significant improvement in inferencing performance, reduction in GPU count, and power reduction – all achieved simultaneously in balance with primary operational metrics.

There are multiple cost and power reduction directions in the \phoenix hardware framework which include a reduction in the number of GPU’s, elimination of local HBM, and the use of lower cost and lower power remote memory. These cost and power optimizations will be offset in small part by the additional cost and power associated with the Tensor Acceleration Bridge. Functional standardization of compute, memory, and interconnects eliminates single source proprietary lock-ins and promotes value chain disaggregation such as GPU to HBM combinations with proprietary GPU to GPU and GPU to CPU interconnects. Decoupling technology transitions between compute, memory, and network with the intent to accelerate new technologies into the inferencing infrastructure without an E2E lock-step revision promotes cost reduction acceleration in the E2E stack.

The discussion on the physical design details is outside the scope of this paper. We expect to bring a detailed write-up on the \phoenix physical design in the next paper in this series.

\chapter{Conclusions}

This section summarizes the results of an extensive evaluation of the \phoenix shared-memory architecture, based on multiple simulation studies, with a focus on its effectiveness in handling diverse AI workloads. 
The simulations show substantial efficiency gains for LLM workloads such as GPT-3, Grok-1, and Qwen-3 235B, demonstrating \phoenix’s capability to reduce total GPU count by up to \textcolor{blue}{\textbf{50\%}}, while simultaneously increasing compute capacity per GPU. 
These results underscore \phoenix’s adaptability and its potential for more efficient resource management in large-scale AI applications.

\bibliographystyle{plain}
\bibliography{refs}

@misc{microsoft_ai_investment_2025,
  author = {Reuters},
  title = {{Microsoft plans to invest \$80 billion on AI-enabled data centers in fiscal 2025}},
  year = {2025},
  howpublished = {\url{https://www.reuters.com/technology/artificial-intelligence/microsoft-plans-spend-80-bln-ai-enabled-data-centers-fiscal-2025-cnbc-reports-2025-01-03/}},
  note         = {Accessed: 2025-07-21}
}

@misc{meta_ai_investment_2025,
  author = {ITPro},
  title = {{Meta is working on a 5GW data center to supercharge AI infrastructure}},
  year = {2025},
  howpublished = {\url{https://www.itpro.com/infrastructure/data-centres/meta-working-on-a-5gw-data-center-to-supercharge-ai-infrastructure-and-mark-zuckerberg-says-one-cluster-alone-covers-a-significant-part-of-the-footprint-of-manhattan}},
  note         = {Accessed: 2025-07-21}
}

@misc{google_ai_investment_2025,
  author = {Reuters},
  title = {{Alphabet CEO reaffirms planned \$75 billion capital spending in 2025}},
  year = {2025},
  howpublished = {\url{https://www.reuters.com/technology/alphabet-ceo-reaffirms-planned-75-billion-capital-spending-2025-2025-04-09/}},
  note         = {Accessed: 2025-07-21}
}

@misc{amazon_ai_investment_2025,
  author = {Rollet, Charles},
  title = {{Amazon doubles down on AI with a massive \$100B spending plan for 2025}},
  year = {2025},
  journal = {TechCrunch},
  howpublished = {\url{https://techcrunch.com/2025/02/06/amazon-doubles-down-on-ai-with-a-massive-100b-spending-plan-for-2025/}},
  note         = {Accessed: 2025-07-21}
}

@misc{explodingtopics2023,
  author       = {Exploding Topics},
  title        = {{ChatGPT Users — Growth Over Time}},
  year         = {2023},
  howpublished = {\url{https://explodingtopics.com/blog/chatgpt-users}},
  note         = {Accessed: 2025-07-21}
}

@misc{thedecoder2023gpt4,
  author       = {The Decoder},
  title        = {{GPT-4 reportedly has 1.76 trillion parameters}},
  year         = {2023},
  howpublished = {\url{https://the-decoder.com/gpt-4-has-a-trillion-parameters/}},
  note         = {Unofficial estimate based on developer insights}
}

@misc{brown2020gpt3,
  author       = {Tom B. Brown et al.},
  title        = {{Language Models are Few-Shot Learners}},
  year         = {2020},
  howpublished = {\url{https://arxiv.org/abs/2005.14165}},
  note         = {GPT-3 with 175B parameters}
}

@misc{smith2021mtnlg,
  author       = {Shaden Smith et al.},
  title        = {{Using DeepSpeed and Megatron to Train Megatron-Turing NLG 530B, the World's Largest and Most Powerful Generative Language Model}},
  year         = {2021},
  howpublished = {\url{https://www.microsoft.com/en-us/research/blog/mt-nlg-530b-a-new-state-of-the-art-model/}},
  note         = {MT-NLG with 530B parameters}
}

@misc{chowdhery2022palm,
  author       = {Aakanksha Chowdhery et al.},
  title        = {{PaLM: Scaling Language Modeling with Pathways}},
  year         = {2022},
  howpublished = {\url{https://arxiv.org/abs/2204.02311}},
  note         = {PaLM with 540B parameters}
}

@misc{fedus2022switch,
  author       = {William Fedus et al.},
  title        = {{Switch Transformers: Scaling to Trillion Parameter Models with Simple and Efficient Sparsity}},
  year         = {2022},
  howpublished = {\url{https://arxiv.org/abs/2101.03961}},
  note         = {Switch-C with 1.6T total parameters (MoE)}
}

@misc{du2022glam,
  author       = {Nan Du et al.},
  title        = {{GLaM: Efficient Scaling of Language Models with Mixture-of-Experts}},
  year         = {2022},
  howpublished = {\url{https://arxiv.org/abs/2112.06905}},
  note         = {GLaM with 1.2T parameters (MoE)}
}

@misc{resourcera2025,
  author       = {Resourcera},
  title        = {{Eye‑Opening AI Statistics (2025): Market Size, Usage \& More}},
  year         = {2025},
  howpublished = {\url{https://resourcera.com/data/artificial-intelligence/ai-statistics/}},
  note         = {Global AI user counts 2020–2025}
}

@misc{altindex2025,
  author       = {AltIndex},
  title        = {{Global AI Adoption to Hit 378 Million Users in 2025}},
  year         = {2025},
  howpublished = {\url{https://altindex.com/news/global-ai-adoption-to-surge}},
  note         = {Forecasted AI user growth through 2030}
}

@techreport{nvidia_h100_whitepaper_2022,
  title        = {{NVIDIA H100 Tensor Core GPU Architecture}},
  author       = {{NVIDIA Corporation}},
  institution  = {NVIDIA},
  type         = {White Paper},
  year         = {2022},
  month        = mar,
  note         = {“Hopper Architecture In‑Depth”},
  url          = {https://www.nvidia.com/hopper-architecture-whitepaper},
}

@techreport{amd_mi300x_whitepaper_2025,
  title        = {{Introducing AMD CDNA3 Architecture: AMD Instinct MI300X and MI300A}},
  author       = {{AMD, Inc.}},
  institution  = {AMD},
  type         = {White Paper},
  year         = {2025},
  month        = jul,
  note         = {Covers architecture, multi-chiplet design, memory/compute connectivity},
  url          = {https://www.amd.com/content/dam/amd/en/documents/instinct-tech-docs/white-papers/amd-cdna-3-white-paper.pdf},
}

@inproceedings{jouppi2023tpuv4,
  author       = {Jouppi, Norman P. and Kurian, George and Li, Sheng and Ma, Peter and Nagarajan, Rahul and Nai, Lifeng and Patil, Nishant and Subramanian, Suvinay and Swing, Andy and Towles, Brian and Young, Cliff and Zhou, Xiang and Zhou, Zongwei and Patterson, David},
  title        = {{TPU v4: An Optically Reconfigurable Supercomputer for Machine Learning with Hardware Support for Embeddings}},
  booktitle    = {Proceedings of the 50th International Symposium on Computer Architecture (ISCA ’23)},
  year         = {2023},
  month        = jun,
  address      = {Orlando, FL, USA},
  pages        = {1--14},
  doi          = {10.1145/3579371.3589350},
  publisher    = {ACM},
}

@manual{nvidia_h200_datasheet_2023,
  title        = {{NVIDIA H200 Tensor Core GPU Datasheet}},
  author       = {{NVIDIA Corporation}},
  organization = {NVIDIA},
  address      = {Santa Clara, CA, USA},
  year         = {2023},
  month        = nov,
  type         = {Datasheet},
  note         = {141GB HBM3e, 4.8TB/s bandwidth; preliminary specifications},
  url          = {https://.../hpc-datasheet-sc23-h200-datasheet-3002446.pdf},
  howpublished = {PDF download},
}

@INPROCEEDINGS{224gserdes,
  author={Pfaff, Dirk and Nummer, Muhammad and Hai, Noman and Xia, Peter and Yang, Kai Ge and Mohsenpour, Mohammad-Mahdi and LaCroix, Marc-Andre and Zamanlooy, Babak and Eeckelaert, Tom and Petrov, Dmitry and Haroun, Mostafa and Dick, Carson and Zaman, Alif and Mei, Haitao and Moazzeni, Shahab and Shakir, Tahseen and Carvalho, Carlos and Huang, Howard and Kumari, Pratibha and Mason, Ralph and Brishty, Fahmida and Jaffri, Ifrah},
  booktitle={2024 IEEE International Solid-State Circuits Conference (ISSCC)}, 
  title={{7.3 A 224Gb/s 3pJ/b 40dB Insertion Loss Transceiver in 3nm FinFET CMOS}}, 
  year={2024},
  volume={67},
  number={},
  pages={128-130},
  doi={10.1109/ISSCC49657.2024.10454537}}

@misc{nvidia2023gracehopper,
  author       = {{NVIDIA}},
  title        = {{NVIDIA GH200 Grace Hopper Superchip Architecture}},
  year         = {2023},
  howpublished = {\url{https://resources.nvidia.com/en-us-grace-cpu/nvidia-grace-hopper}},
  note         = {Accessed: 2025-07-23}
}

@misc{nvidia2024gb200nvl72,
  author       = {{NVIDIA}},
  title        = {{NVIDIA GB200 NVL72 Architecture}},
  year         = {2024},
  howpublished = {\url{https://www.nvidia.com/en-us/data-center/gb200-nvl72/}},
  note         = {Accessed: 2025-07-23}
}

@misc{nvidia2024blackwell,
  author       = {{NVIDIA}},
  title        = {{NVIDIA Blackwell Architecture Technical Overview}},
  year         = {2024},
  howpublished = {\url{https://resources.nvidia.com/en-us-blackwell-architecture}},
  note         = {Accessed: 2025-07-23}
}

@misc{kaist2025hbmroadmap,
  author       = {{TeraByte Interconnection and Package Laboratory, KAIST}},
  title        = {{2025 Teralab Next-Gen. HBM Roadmap}},
  year         = {2025},
  howpublished = {\url{https://tera.kaist.ac.kr/researches/teralab-hbm-milestone-and-roadmap}},
  note         = {Accessed: 2025-11-8}
}

@INPROCEEDINGS {google2025hotchip,
author = { Jouppi, Norman P. and Lakshmanamurthy, Sridhar },
booktitle = { 2025 IEEE Hot Chips 37 Symposium (HCS) },
title = {{ Ironwood: Delivering Best in Class perf, perf/TCO and perf/Watt for Reasoning Model Training and Serving }},
year = {2025},
volume = {},
ISSN = {},
pages = {1-26},
doi = {10.1109/HCS66204.2025.11154400},
url = {https://doi.ieeecomputersociety.org/10.1109/HCS66204.2025.11154400},
publisher = {IEEE Computer Society},
address = {Los Alamitos, CA, USA},
month =Aug}

@misc{intel2025aichip,
  author       = {Intel},
  title        = {{Intel to Expand AI Accelerator Portfolio with New GPU}},
  year         = {2025},
  howpublished = {\url{https://newsroom.intel.com/artificial-intelligence/intel-to-expand-ai-accelerator-portfolio-with-new-gpu}},
  note         = {Accessed: 2025-11-8}
}

@misc{nvidia2025davies,
      title={{Efficient LLM Inference: Bandwidth, Compute, Synchronization, and Capacity are all you need}}, 
      author={Michael Davies and Neal Crago and Karthikeyan Sankaralingam and Christos Kozyrakis},
      year={2025},
      eprint={2507.14397},
      archivePrefix={arXiv},
      primaryClass={cs.AR},
      url={https://arxiv.org/abs/2507.14397}, 
}

@ARTICLE{gatech20253ddram,
  author={Hsu, Po-Kai and Sharda, Janak and Wu, Xiangjin and Wong, H.-S. Philip and Yu, Shimeng},
  journal={IEEE Nanotechnology Magazine}, 
  title={{Monolithic 3D Stackable DRAM}}, 
  year={2025},
  volume={19},
  number={2},
  pages={7-16},
  doi={10.1109/MNANO.2025.3533815}}

@misc{nvidia-ai-factory,
  author       = {NVIDIA},
  title        = {{Data Center Solutions: AI Factories}},
  year         = {2025},
  howpublished = {\url{https://www.nvidia.com/en-us/solutions/ai-factories/}},
  note         = {Accessed: 2025-11-8}
}

@misc{blog-ai-factory,
  author       = {NVIDIA},
  title        = {{AI Factories Are Redefining Data Centers and Enabling the Next Era of AI}},
  year         = {2025},
  howpublished = {\url{https://blogs.nvidia.com/blog/ai-factory/}},
  note         = {Accessed: 2025-11-8}
}

\end{document}